\documentstyle[epsf,12pt]{article}

\catcode`@=11 \@addtoreset{equation}{section} \catcode`@=12

\def\strutdepth{\dp\strutbox}
\def\marginarrow#1{\vtop to\strutdepth{
    \baselineskip\strutdepth\vss\llap{#1$\rightarrow$ }\null}}
\def\revision#1{\strut\vadjust{\kern-\strutdepth\marginarrow{#1}}}

\begin{document}
\begin{titlepage}
\begin{flushright}\vbox{\begin{tabular}{c}
           TIFR/TH/98-23\\
           June, 1998\\
           hep-lat/9806034\\
\end{tabular}}\end{flushright}
\begin{center}
   {\large \bf
      Dimensional Reduction and Screening Masses\\
      in Pure Gauge Theories at Finite Temperature}
\end{center}
\bigskip
\begin{center}
   {Saumen Datta\footnote{E-mail: saumen@theory.tifr.res.in} and
    Sourendu Gupta\footnote{E-mail: sgupta@theory.tifr.res.in}\\
    Department of Theoretical Physics,\\
    Tata Institute of Fundamental Research,\\
    Homi Bhabha Road, Mumbai 400005, India.}
\end{center}
\bigskip
\begin{abstract}
We studied screening masses in the equilibrium thermodynamics of $SU(2)$ and
$SU(3)$ pure gauge theories on the lattice. At a temperature of $2T_c$ we
found strong evidence for dimensional reduction in the non-perturbative
spectrum of screening masses. Mass ratios in the high temperature $SU(3)$
theory are consistent with those in the pure gauge theory in three dimensions.
At the first order $SU(3)$ phase transition we report the first measurement
of the true scalar screening mass.
\end{abstract}
\end{titlepage}

\section{\label{sc.intro}Introduction}

The equilibrium thermodynamics of a gauge theory is studied in the
non-perturbative domain by lattice simulations of the partition
function. Much is now known about the phase transitions in $SU(2)$
and $SU(3)$ pure gauge theory and in QCD, including the order, the
transition temperature, $T_c$, entropy density, pressure, specific
and latent heats and other such quantities \cite{rev}. 

Also of interest are the screening masses at finite temperatures.
These are defined in general by the exponential spatial falloff of
the correlation of two static sources. A classification of all masses
is provided by the transformation properties of the sources. For
glueball-like screening masses, such a classification was
performed in \cite{old} where the first measurements of several of
these masses were reported. Extensive lattice data is available on
screening masses from meson and baryon-like sources \cite{dyn}, and
on the correlation of Polyakov lines.

A gauge invariant transfer matrix formulation is easy to write for the
lattice regularised theory. For thermal physics, it is convenient to
think of the transfer matrix in a spatial direction. The free energy and
other bulk thermodynamic quantities involve only the largest eigenvalue
of the transfer matrix. Correlation functions and the screening masses
involve the ratio of this largest eigenvalue with specific other eigenvalues
depending on the transformation properties of the source. Thus, the
full spectrum of screening masses contains much more information about
the theory than bulk thermodynamics can provide.

A quantity of special interest is the Debye screening mass, $M_D$, whose
inverse gives the screening length of static choromo-electric fields. In
the quantum theory, this screening mass plays an important role in
regulating some infra-red singularities. Lattice measurements have, in
the past, concentrated on measuring a mass from the correlation of Polyakov
loops, $M_P$. It was expected that when the gauge coupling becomes small,
$M_P=2M_D$. Many years ago Nadkarni showed that the relation between $M_D$
and $M_P$ is far from being so simple \cite{shirish}. Subsequent lattice
work focussed on methods of computing $M_D$.

Reisz and collaborators \cite{thomas,pierre} wrote down a dimensionally
reduced theory which could be used to define $M_D$ non-perturbatively. A
recent paper \cite{yaffe} used the representations of the symmetries of
the transfer matrix to write down operators whose correlations could be
used to measure the Debye mass\footnote{The group theoretical identification
of $M_P$ and $M_D$ was, in fact, first given in \cite{old}.} and gave a
general parametrisation of the perturbative and non-perturbative terms for
$M_D$. These parameters have since been determined in a lattice measurement
of the Debye screening mass using a dimensionally reduced theory at very
high temperatures \cite{kari1}.

A similar screening mass is required for the magnetic sector of the
nonabelian gauge theory in the plasma, in order to get sensible
results in the infrared. However, the magnetic mass is entirely
non-perturbative in nature. It has been the object of many lattice studies
\cite{maglat,rank}. In fact, a recent work \cite{rank} tries to extract
$M_D$ and $M_m$ from gauge fixed gluon propagators at finite temperature.

Many years ago Linde discovered \cite{linde} that the thermal
perturbation expansion in non-Abelian gauge theories breaks down
because the magnetic sector is not amenable to perturbative studies. 
A recent attempt to understand Linde's problem in the region where the
coupling $g\ll1$ has invoked a sequence of dimensionally reduced effective
theories \cite{braaten}. At length scales of $1/gT$ dimensional reduction
yields a three dimensional $SU(N)$ gauge theory coupled to an adjoint scalar
field of mass $M_D\approx{\cal O}(gT)$. At longer scales, $1/g^2T$, the
scalar field can be integrated out, and the leading terms in the effective
theory correspond to a pure gauge theory in three dimensions. On the
basis of this reduction it has been argued \cite{braaten,yaffe} that a
non-vanishing pole in magnetic gluon propagators is absent, and
Linde's problem is finessed by confinement in the three dimensional
gauge theory.

In fact, dimensional reduction has often been used to explore finite
temperature theories \cite{dimred}, and it has long been argued that the
dimensionally reduced theory is confining. The spatial string tension is
known to be non-vanishing for $T>T_c$ and scales as $T^2$ \cite{spstr}.
This, and other, lattice measurements reveal that the gauge coupling close
to $T_c$ are too large to trust perturbation theory.

At temperatures of a few $T_c$, the coupling $g\ge1$, and the length
scales $1/T\approx1/gT\approx1/gT^2$, and hence cannot be decoupled. The
construction of \cite{braaten} is not applicable. However, the transfer
matrix argument assures us that $M_D$ is only the smallest in a hierarchy
of screening masses. We can then use the spectrum of the screening masses
to explore the symmetries of the transfer matrix and check whether or not
dimensional reduction works; and if it does, then what is the nature of the
reduced theory.

Our major result is that the degeneracies
of the spectrum of screening masses implies that the symmetry group of the
spatial transfer matrix is that of two dimensional rotations--- implying
dimensional reduction for $T\approx2T_c$. The measured non-perturbative
spectrum is similar to that in 3-d pure gauge theory \cite{2plus1}.

The symmetries of the transfer matrix, and the physical consequences are
discussed in Section \ref{sc.group}. The group theory presented in this
section is central to the rest of this paper. The extraction of screening
masses in $T>0$ four-dimensional $SU(3)$ and $SU(2)$ pure gauge theories
take up the next two sections. These may be skipped by those readers who
are not interested in the details of the lattice simulations. Section
\ref{sc.final} contains a summary of our lattice results and presents
our conclusions on the nature of the dimensionally reduced theory. Several
technical details are relegated to appendices. Appendix \ref{sc.irreps} 
gives the loop operators used in our computations. In Appendix 
\ref{sc.clebsch}, we examine the possibility of explaining the mass 
spectrum in terms of perturbative multigluon states. Noise reduction 
techniques for the lattice simulations and the algorithm for projecting 
on to the lowest state in every channel are described in Appendices 
\ref{sc.fuzz} and \ref{sc.vary} respectively.

\section{\label{sc.group}Symmetries of the Transfer Matrix}

\subsection{Group chains}

For the $T=0$ continuum Euclidean theory the symmetry of the transfer
matrix is the direct product of the full rotation group $O(3)$ and the
$Z_2$ groups generated by charge conjugation, $C$, and time reversal.
Irreps of $O(3)$ are labelled by the angular momentum and parity, $J^P$,
and of the full symmetry group by $J^{PC}$. For the lattice regularised
theory, $O(3)$ breaks to the discrete subgroup of the symmetries of the
cube, $O_h$. The consequent reduction of the irreps of $O(3)$ is
well-known \cite{bb}.

In the Euclidean formulation of the (continuum) equilibrium $T>0$ theory,
the transfer matrix in one of the spatial directions is invariant under
symmetries of the orthogonal slice. Such slices are three dimensional---
two of which are spatial and one is the Euclidean time. The symmetry
group is that of a cylinder, ${\cal C}=O(2)\times Z_2$. This $Z_2$
factor is generated by $\sigma_z\;:\;t\mapsto-t$. The non-Abelian group
$O(2)$ contains the Abelian subgroup of rotations, $SO(2)$, and a
2-d parity, $\Pi\;:\;(x,y)\mapsto(x,-y)$. The 3-d parity $P=C(\pi)\sigma_z$,
where $C(\pi)$ is the rotation by $\pi$ in $SO(2)$.

In the high temperature limit, the effective theory is expected to undergo
dimensional reduction to a 3-d gauge theory. In such a theory, the transfer
matrix must have the $O(2)$ symmetry of a 2-d slice. $O(2)$ has two
one-dimensional irreps $0_\pm$ (the 2-d scalar and pseudo-scalar) and an
infinite tower of two-dimensional irreps $M$. Under the reduction of $SO(3)$
to $O(2)$, the spin $J$ irrep breaks as---
\begin{equation}
   J\;\to\; 0_{\Pi(J)} + \sum_{M=1}^J M,\qquad
   {\rm where}\quad \Pi(J)=(-1)^J.
\label{o2}\end{equation}

On the lattice the irreps of $\cal C$ break further into irreps of the
automorphism group of a $z$-slice, the tetragonal group $D^4_h=D_4\times
Z_2(P)$. For the three dimensional lattice theory, the automorphism
group of the 2-d slice is $C^4_v$, which is isomorphic to $D_4$. A recent
work on 3-d glueballs used this classification of the states \cite{2plus1}.

This pattern of symmetry breaking is summarised by
\begin{equation}
 \matrix{
     O(3)=SO(3)\times Z_2(P) &\longrightarrow& O_h=O\times Z_2(P)\cr
     \bigg\downarrow & & \bigg\downarrow \cr
     {\cal C}=O(2)\times Z_2(\sigma_z) &\longrightarrow&
                         D^4_h=D^4\times Z_2(P)\cr
     \bigg\downarrow & & \bigg\downarrow \cr
     O(2) &\longrightarrow& C^4_v\cr
        }
\label{redux}\end{equation}
Four of the $Z_2$ factor groups are identical, and generated by the 3-d parity
$P$. $O(2)$ and $C^4_v$ contain the 2-d parity $\Pi$.

\subsection{Point group representations}

In this paper we use the notation of \cite{hammer} for the crystallographic
point groups and their irreducible representations (irreps). This subsection
contains a discussion of the irreducible representations of the lattice
symmetries.

The group $D_4$ is generated by eight elements in five conjugacy classes--- the
identity ($E$), rotations of $\pm\pi/2$ around the $z$-axis ($C_4$), rotations
of $\pi$ around the $z$-axis ($C_4^2$), around the $x$ or $y$-axes ($C_2$)
and around the directions $x\pm y$ ($C_2'=C_4C_2$). The five irreps are
called $A_1$, $A_2$, $B_1$, $B_2$ and $E$. The first four are one-dimensional
and the last two-dimensional. The 10 irreps of $D^4_h$ are obtained by
adjoining the character of $P$ to the irreps of $D_4$. Under the reduction
of $O_h$ to $D^4_h$, the irreps break up as---
\begin{eqnarray}
   A_1^P\;&\to&\;A_1^P,\;\;\qquad\qquad A_2^P\;\to\;B_1^P,\cr
   T_1^P\;&\to&\;A_2^P + E^P,\qquad T_2^P\;\to\;B_2^P + E^P,\cr
   E^P\;&\to&\;A_1^P + B_1^P.
\label{ohtod4h}\end{eqnarray}
The notation $A_{1,2}^\pm$ is potentially confusing since these irreps can
belong to both $O_h$ and $D^4_h$. A similar confusion can also be caused by
the notation $E$, which stands for the two-dimensional irrep of any group.
In the rest of this paper we will mention the group involved whenever we
mention one of these irreps.

In the dimensionally reduced theory, the lattice symmetry group is $C^4_v
\simeq D_4$. Instead of calling the irreps by the names of the $D_4$ irreps,
we denote them by the symbols $A^+$, $A^-$, $B^+$, $B^-$ and $E$ respectively.
The first four are one-dimensional irreps, and the sign which indexes them is
the character of the 2-d parity $\Pi$ in the irrep. The breaking of $D^4_h$
irreps to $C^4_v$ is as---
\begin{eqnarray}
   A_1^+,A_2^-\;&\to&\;A^+,\qquad A_1^-,A_2^+\;\to\;A^-,\cr
   B_1^+,B_2^-\;&\to&\;B^+,\qquad B_1^-,B_2^+\;\to\;B^-,\cr
   E^P\;\;&\to&\;E.\qquad
\label{d4htoc4}\end{eqnarray}
Our nomenclature for the irreps of $D^4_h$ is designed to be a mnemonic for
the breaking of the $O_h$ irreps (eq.\ \ref{ohtod4h}), but not for the
breaking of $D^4_h$ to $C^4_v$ (eq.\ \ref{d4htoc4}). On the other hand, the
naming convention followed in \cite{old,yaffe} simplifies the latter at the
expense of the former, since it proceeds from the isomorphism $D^4_h\simeq
C^4_v\times Z_2(\sigma_z)$. To make contact with the notation used elsewhere,
note that $P$ in \cite{yaffe} corresponds to our 2-d parity $\Pi$ and
$R_{\bar z}$ to the character of the operator $\sigma_z$, which is called $P$
in \cite{old}.

The $O(2)$ irreps break to $C^4_v$ as follows---
\begin{eqnarray}
   0_+\;&\to&\;A^+,\qquad\qquad{\rm and}\qquad0_-\;\to\;A^-,\cr
   M\;&\to&\;
     \cases{
        E & $(M={\rm odd}),\qquad$\cr
        B^+ + B^- & $(M=2 {\rm\ mod\ }4),$\cr
        A^+ + A^- & $(M=0 {\rm\ mod\ }4).$\cr
           }
\label{ctod4h}\end{eqnarray}
The $A^+$ corresponds to the scalar and the $A^-$ to the pseudo-scalar
irrep of $O(2)$.

States or operators also carry a label $C=\pm1$ for the charge conjugation
symmetry. The charge conjugation operator reverses the direction of traversal
of a loop, and hence takes the trace into its complex conjugate. $C=1$ states
correspond to real parts of loops and $C=-1$ states to imaginary parts.
For gauge groups with only real representations, such as $SU(2)$, there
are no $C=-1$ states. The construction of irreps of $D^4_h$ from loops is
given in Appendix \ref{sc.irreps}.

\begin{figure}
\vskip6truecm
\includegraphics{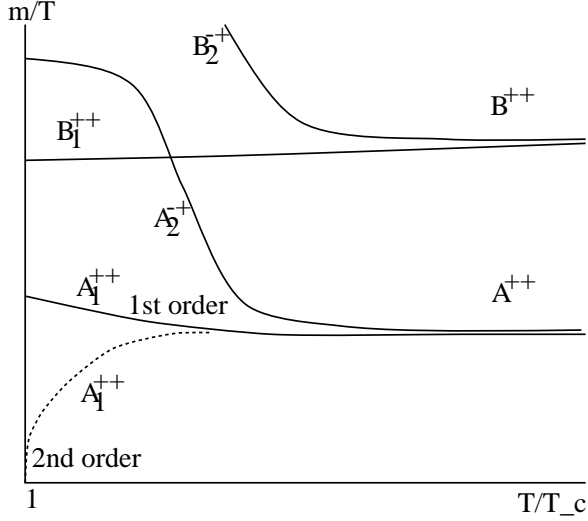}
\caption[dummy]{Possible temperature dependence of the lowest screening
    masses in some channels. Dimensional reduction at high temperatures
    is signalled by a pattern of approximate degeneracies characteristic
    of $C^4_v$ symmetry.}
\label{fg.guess}\end{figure}

In finite temperature lattice simulations, the symmetry group of the
transfer matrix is always $D^4_h$. However, at low temperatures, we should
expect to see an effective symmetry group $O_h$. At the other end of the
temperature scale, if dimensional reduction is to be a good approximation,
we should see an approximate $C^4_v$ symmetry. A guess at the temperature
dependence of the screening masses is shown in Figure \ref{fg.guess}. If
the lattice is big enough, and the lattice spacing is sufficiently small,
then we should see the more extended degeneracy of $O(2)$. In this case
the $A^+$ and $A^-$ of $C^4_v$ will be non-degenerate, since they
correspond to different irreps of $O(2)$, but the $B^+$ and $B^-$ will
become degenerate, since they correspond to one irrep of $O(2)$.
Differences between first and second order phase transitions should be
visible in the $A_1^{++}$ channel near $T_c$.

\section{\label{sc.su3}$SU(3)$ Pure Gauge Theory}

We have simulated the $SU(3)$ pure gauge theory with Wilson action
at three temperatures. With $N_\tau=4$ the critical coupling
is $\beta_c(N_\tau=4)\;=\;5.692$ \cite{4su3, 6su3}. We have performed a
simulation at $T\approx T_c$ with $\beta=5.7$. At $T=3T_c/2$ the
coupling is $\beta_c(N_\tau=6)\approx5.9$ \cite{6su3}, and for $T=2T_c$
we use the coupling $\beta_c(N_\tau=8)\approx6.0$ \cite{nsu3}.

The simulations were performed with a Cabbibo-Marinari pseudo heat-bath
update, where each update acted on three separate $SU(2)$ subgroups by
five Kennedy-Pendleton moves \cite{kphb}. The class of loop operators measured
is listed in Appendix \ref{sc.irreps}. Noise reduction involved a fuzzing
procedure explained in Appendix \ref{sc.fuzz}. For each irrep of the symmetry
group, we projected the measured correlation function to the ground state by
a variational technique explained in Appendix \ref{sc.vary}. Successive
measurements of correlation functions were separated by about one integrated
auto-correlation time measured through the Polyakov loop.

One of the techniques for extracting screening masses from correlation
functions in the long direction (with $N_z$ sites) is to solve the equation
\begin{equation}
   {\cosh\left[m(z+1/2)(N_z/2-z-1)\right]\over
                \cosh\left[m(z+1/2)(N_z/2-z)\right]}
     \;=\; {C(z+1)\over C(z)}
\label{su3.local}\end{equation}
for the local mass, $m(z+1/2)$, given measurements of the correlation
function $C(z+1)$ and $C(z)$. The assumption that the correlation function
is described by a single mass is borne out if there is a range of $z$ for
which the local mass is constant within errors. If there is such a plateau
then we quote it as our estimate of the screening mass.

In addition we have performed fits to correlation functions in the form
\begin{equation}
   C(z)\;=\;C(0)\left\{r{\cosh\left[\mu_0(N_z/2-z)\right]\over
                           \cosh(\mu_0 N_z/2)}
              +(1-r){\cosh\left[\mu_1(N_z/2-z)\right]\over
                           \cosh(\mu_1 N_z/2)} \right\},
\label{su3.form2}\end{equation}
where $r$, $\mu_0$ and $\mu_1$ are fit parameters with the constraint
$\mu_0<\mu_1$. Local mass estimates were deemed acceptable if the fitted
value of $\mu_0$ agreed with it. When the local masses were too noisy to
show a plateau, we took $\mu_0$ as our estimate of the screening mass.
The goodness of the variational projection to the ground state was checked
by observing whether the fitted parameter $r$ was close to unity.

In minimising $\chi^2$ to perform the fits we took into account covariances
of data \cite{gott}. Since we normalised the variational correlator at
separation zero to unity, the error in this point is zero. However, the
intrinsic variability in the zero distance correlator is then redistributed
over all the points through the covariance. The maximum distance retained in
the fit was always determined by the criterion that the correlation function
at that distance should be more than 1-$\sigma$ from zero. The estimates and
errors of every measurable were constructed through a jack-knife procedure.

\subsection{\label{sc.su3tc}$T=T_c$}

\begin{figure}
\vskip10truecm
\includegraphics{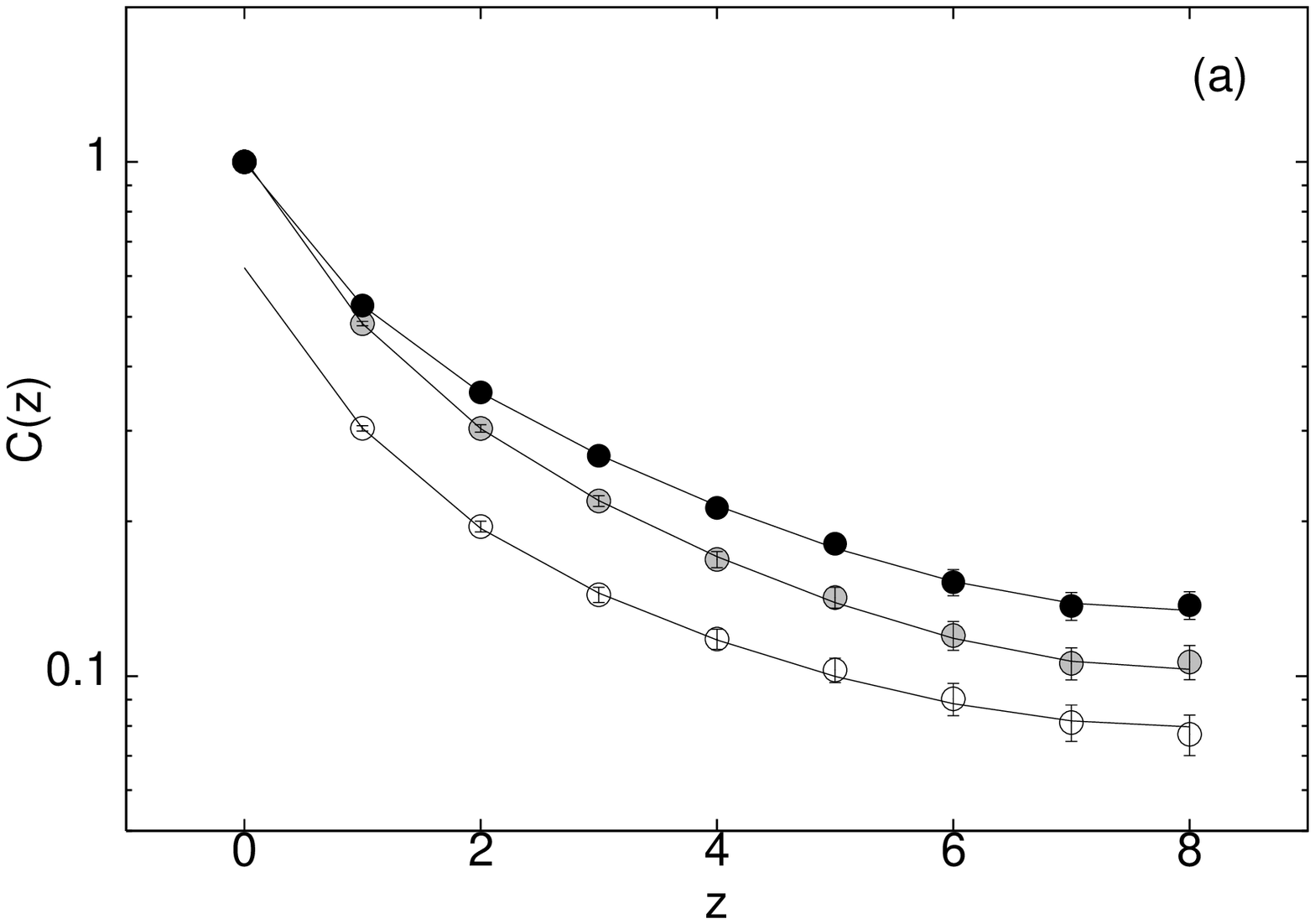}
\includegraphics{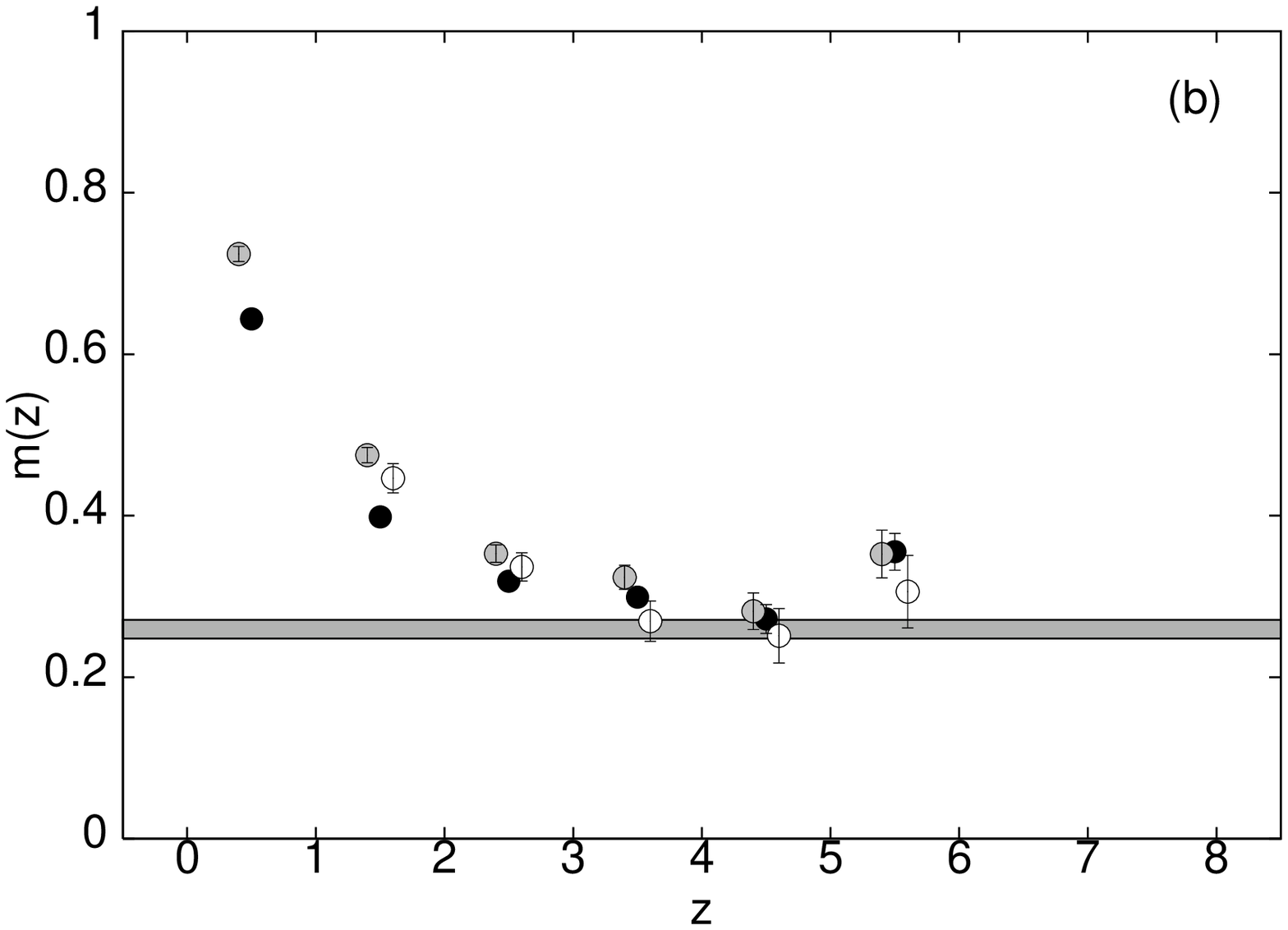}
\includegraphics{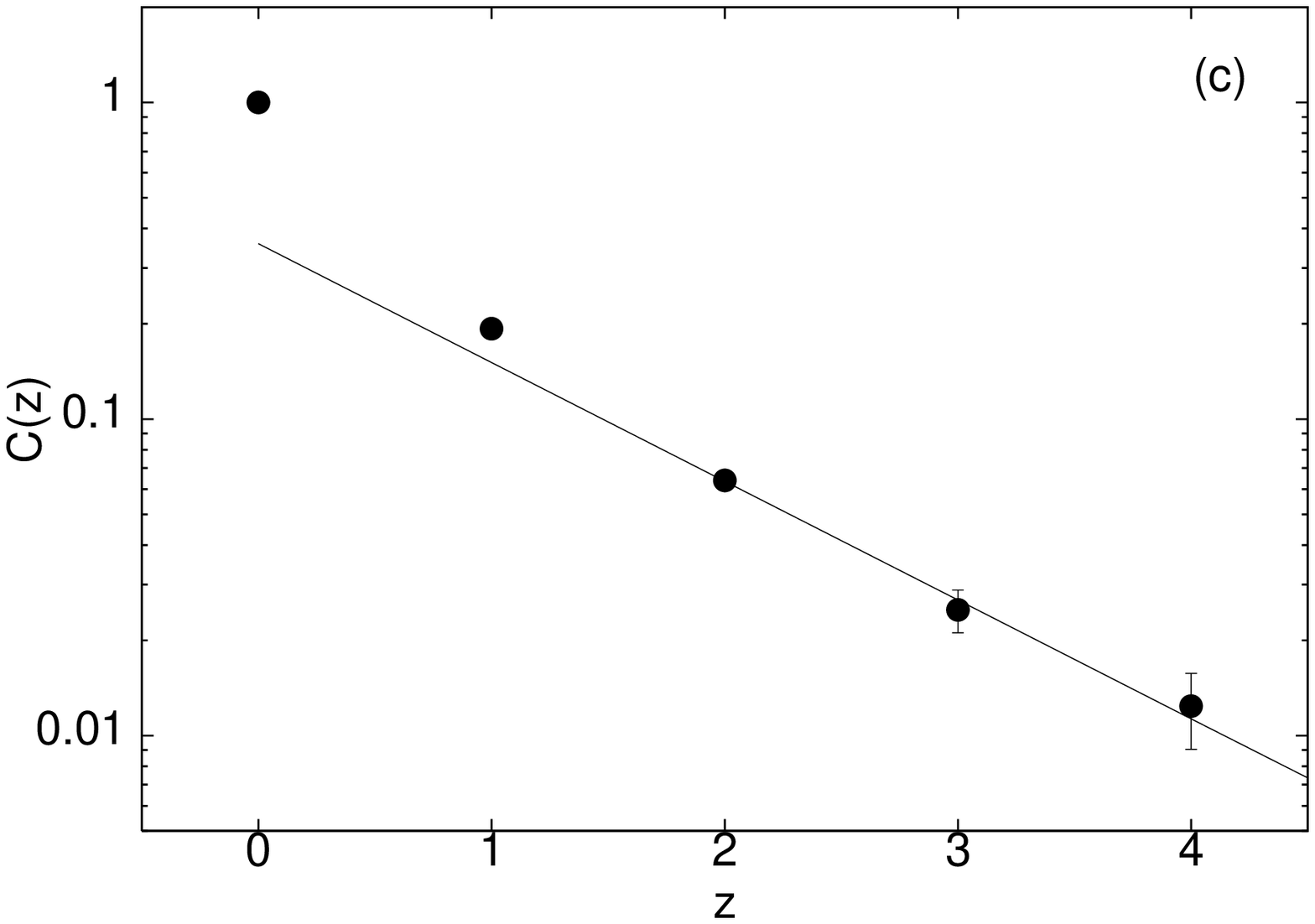}
\includegraphics{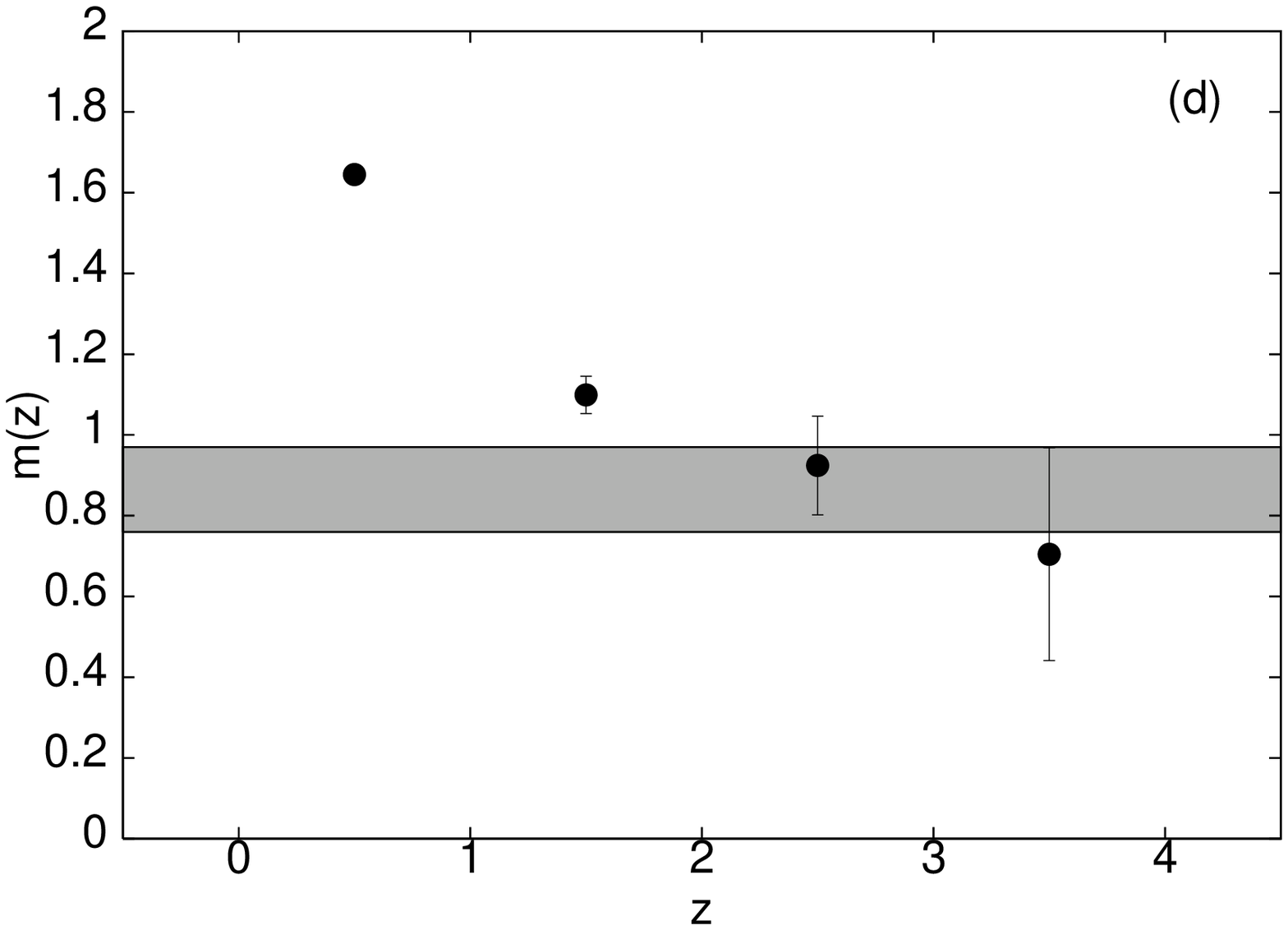}
\caption[dummy]{$A_1^{++}$ correlation functions in $SU(3)$ at $\beta=5.7$
   with $(0,1)$ variation. (a) Data and fits for the projection to the lowest
   mass (thermal $A_1^{++}$: dark circles, $T=0$ $A_1^{++}$: grey circles,
   $T=0$ $E^{++}$: white circles), (b) local masses compared to the 1-$\sigma$
   band around the best fit, (c) data and fits to the projection on the second
   mass, and (d) local masses compared to the 1-$\sigma$ error band on the
   fitted physical $A_1^{++}$ mass.}
\label{fg.therm1}\end{figure}

\begin{table}[hbt]\begin{center}
  \begin{tabular}{|c|c|c|c|c|c|c|}  \hline
  Operator & \multicolumn{3}{c|}{$(0,1)$} & \multicolumn{3}{c|}{$(0,2)$}\\
  \cline{2-7} 
  & $\chi^2$ & $r$ & $\mu_0$ & $\chi^2$ & $r$ & $\mu_0$ \\ 
  \hline
  $\left.A_1^{++}\right|_0$&
             $5.7/6$  & $0.71_{-0.05}^{+0.04}$ & $0.27_{-0.02}^{+0.01}$ &
             $6.3/6$  & $0.74_{-0.05}^{+0.04}$ & $0.27_{-0.02}^{+0.01}$ \\
  $\left.E^{++}\right|_0$&
             $2.6/6$  & $0.68_{-0.07}^{+0.05}$ & $0.23_{-0.02}^{+0.02}$ &
             $2.0/6$  & $0.71_{-0.07}^{+0.05}$ & $0.23_{-0.02}^{+0.02}$ \\
  $\left.A_1^{++}\right|_T$&
             $10.5/6$ & $0.81_{-0.04}^{+0.03}$ & $0.26_{-0.01}^{+0.01}$ &
             $10.9/6$ & $0.83_{-0.04}^{+0.03}$ & $0.26_{-0.01}^{+0.01}$ \\
  $\left.B_1^{++}\right|_T$&
                      & & $1.23\pm0.08$ &
                      & & $1.2 \pm0.1 $ \\
  \hline
  \end{tabular}\end{center}
  \caption[dummy]{Fitted masses in $SU(3)$ theory for $\beta=5.7$ on the
     $4\times8^2\times16$ lattice using 5000 configurations. Note that
     $r$ depends on the basis set of operators, but $\mu_0$ does not.
     For the $B_1^{++}$ operator, a plateau in the local masses is
     used for the estimate of the screening mass.}
\label{tb.tc}\end{table}

At $\beta=5.7$ we analysed configurations separated by 50 pseudo-heat-bath
sweeps, discarding the first 20 configurations for thermalisation. Tunnelling
between phases occurred every 600 sweeps on an average. The operators measured
are listed in Appendix \ref{sc.irreps}. Analyses were performed in subspaces
corresponding to the $O_h$ and $D^4_h$ irreps.

From the results in Table \ref{tb.tc}, it is clear that the lowest screening
mass observed in the $\left.A_1^{++}\right|_0$ and $\left.E^{++}\right|_0$
sectors coincides with that obtained in the $\left.A_1^{++}\right|_T$
sector. This is also clear from Figure \ref{fg.therm1}. In addition, the
eigenvector of the variation over all operators was orthogonal (within
errors) to the $\left.B_1^{++}\right|_T$ space and yielded the same mass
as the $\left.A_1^{++}\right|_T$. In the critical region, therefore, the
spectrum of screening masses is organised in irreps of $D^4_h$. In \cite{old}
it was found that the screening masses could be organised into irreps of
$D^4_h$ at $3T_c/2$, but not at $T_c/2$. Our observation extends this to
the picture presented in Figure \ref{fg.guess}.

It is interesting to note that the projection to the $A_1^{++}$ ground
state, as measured by $r$, increases with the number of operators used.
This is a generic feature of the variational method and clearly seen in
Figure \ref{fg.therm1}b. Another generic feature is visible in the
same figure--- the fitted mass is the same as the stable long-distance
local mass. Also, because the fit uses all the data points, its error is
slightly smaller than that of the local mass.

The lowest screening mass in the $\left.A_1^{++}\right|_T$ sector agrees
with previous estimates of the mass from Polyakov line correlations
\cite{brown}. This is the ``tunnelling mass'', which goes to zero and
decouples from the thermodynamics in the infinite volume limit. The mass
in this channel, in the thermodynamic limit, should be finite, and on any
finite lattice it would be the next-to-lowest $A_1^{++}$ mass. We attempted
to estimate this ``physical mass'' by solving the variational problem for
the second lowest eigenvalue. A common estimate of the mass is obtained
from variation over all operators as well as only the $\left.A_1^{++}
\right|_T$ operators. This indicates that the ``physical mass'' in the
$\left.A_1^{++}\right|_T$ sector is genuinely the lowest screening mass.
Local masses and fits agree (see Figure \ref{fg.therm1}) and give
a physical mass
\begin{equation}
   am(A_1^{++})\;=\;0.86_{-0.10}^{+0.11},
    \qquad\qquad(N_\tau=4,\;\;\beta=5.7).
\end{equation}
Such a physical mass has also been estimated before for two 3-state spin
models in three dimensions. An $SU(3)$ spin model, obtainable from $SU(3)$
gauge theory in the strong coupling regime, had physical mass
$0.71\pm0.06$ \cite{z3spin} whereas the three state Potts model gave
a mass of about $0.1$ \cite{potts3}. The $A_1^{++}$ mass measured at $T=0$
at $\beta_c=5.7$ is $0.964\pm0.012$ \cite{chen}, giving
\begin{equation}
   {m(T_c,A_1^{++})\over m(T=0,A_1^{++})}\;=\;0.89\pm0.10.
\label{ratc}\end{equation}
In the scaling limit this ratio should be independent of $\beta_c$.

The $\left.B_1^{++}\right|_T$ operator is always a difference of loops.
As a result the correlation function in this sector is much more noisy
than in the $A_1^{++}$ sector. Nevertheless, we were able to follow the
correlation function to distance 4, and obtain a plateau in the local
masses. Our estimate of the screening mass, reported in Table \ref{tb.tc},
comes from the local mass, $m(5/2)$ for both $(0,1)$ and $(0,2)$ variation.

\subsection{\label{sc.su3hi}$T=3T_c/2$ and $2T_c$}

We made runs on $4\times8^2\times16$ lattices at $\beta=5.9$ and 6.1,
corresponding to $3T_c/2$ and $2T_c$ respectively. Since the integrated
autocorrelation time for the Polyakov loop was estimated to be less than
10 sweeps, we analysed data separated by 10 sweeps after discarding the
first 400 for thermalisation. 5000 configurations were generated at
$3T_c/2$ and 10000 at $2T_c$. At $2T_c$ we made two further runs. One was
on a larger, $4\times12^2 \times16$ lattice, where we collected 5000
configurations separated by 10 sweeps, after discarding the first 400
sweeps. The second was on a shorter $4\times8^2\times12$ lattice, with
exactly the same statistics. We did not see any tunnelling events at all
in any of the four runs; from a hot start the system quickly relaxed into
one of the $Z_3$ symmetric free-energy minima,
and stayed there for the duration of the runs. Since measurements of
autocorrelations of correlation functions showed that the integrated
autocorrelation time did not exceed 1.5 measurements, these contributions
to error estimates have been neglected in this section.

We made measurements of the operators listed in Appendix \ref{sc.irreps}.
Each operator was replicated at five levels of fuzzing. In most channels
we could follow the correlation functions out to distance 5 and found the
$(0,z)$ variational ground state to be statistically well behaved even
with $z$ as large as 3. The exceptions were the $B_1^{-+}$ and $B_2^{++}$
channels, which could be followed only to distance 3. As a result the
variational ground state was stable only for $z=1$ and 2.

In each channel, the components of $|0;z\rangle_T$, the $(0,z)$ variational
ground state at temperature $T$, give the overlaps of the ground state with
each operator. We normalised $|0;z\rangle_T$ to unity in each jack-knife bin.
We found that several components of this vector are numerically very stable
from one jack-knife bin to another. The rest of the components fluctuate
from bin to bin, and seem to fine tune the variational eigenvalue. The
eigenvalue itself is far more stable than any of the eigenvectors.

We found that correlation functions at distance 1 differed qualitatively from
the long distance correlation function in several respects, thus giving rise
to certain systematics in the measurement of screening masses.
\begin{itemize}
\item
   The overlap $\langle0;1|0;3\rangle_{2T_c}$ differed significantly from
   unity. For example, this overlap was $0.73\pm0.02$ in the $A_1^{++}$
   channel and $0.57\pm0.05$ for the $A_2^{-+}$. On the other hand, the
   overlap $\langle0;2|0;3\rangle_{2T_c}=0.90\pm0.06$, consistent with unity,
   for $A_1^{++}$. Similar results were obtained at $3T_c/2$.
\item
   There was a strong effect on local masses. With $(0,1)$ variation, we
   usually found no plateau in the local masses. With $(0,2)$ or $(0,3)$
   variations a plateau was often visible.
\item
   Fits to correlation functions also reflected this behaviour. No acceptable
   fit with one or two masses was found to the $(0,1)$ correlation function,
   whereas the $(0,3)$ correlator could be fitted with $r\simeq1$.
\end{itemize}
In view of this, we quote results from the $(0,3)$ variational correlators
in all channels except the $B_1^{-+}$ and $B_2^{++}$, for which we quote
results from $(0,2)$ variation. All the screening masses we quote are
obtained from local masses and verified by a fit. The exception is the
$A_2^{-+}$ channel which turns out to be noisy. In this case the quoted
result is the fitted mass.

The ground states in the $A_1^{-+}$, $A_2^{-+}$ and $B_2^{-+}$ channels
are temperature independent within errors. This is not so in the $A_1^{++}$
and $B_1^{++}$ channels; the overlap of these two ground states at $3T_c/2$
and $2T_c$ are both $0.58\pm0.02$. Interestingly, the masses are independent
of the temperature.

\begin{table}[hbt]\begin{center}
  \begin{tabular}{|c|c|c|c|c|c|}
  \hline
  $O(2)$ & $D^4_h$ & $3T_c/2$ & \multicolumn{3}{c|}{$2T_c$} \\
  \cline{4-6}
  irreps & irreps  & small & short & small & large \\
  \hline
  $0_+^+$ & $A_1^{++}$ & $0.64\pm0.01$   
          & $0.64\pm0.01$   & $0.65\pm0.01$   & $0.65\pm0.02$ \\
  $0_+^+$ & $A_2^{-+}$ & $0.80\pm0.02^*$ 
          & $0.61\pm0.07^*$ & $0.69\pm0.04^*$ & $0.68\pm0.04^*$ \\
  $0_-^+$ & $A_1^{-+}$ & $1.58\pm0.03$   
          & $1.61\pm0.06$   & $1.58\pm0.04$   & $1.56\pm0.04$ \\
  $2^+$   & $B_1^{++}$ & $1.22\pm0.08$   
          & $1.22\pm0.05$   & $1.2\pm0.1$     & $1.27\pm0.06$ \\
  $2^+$   & $B_1^{-+}$ &
          & $1.2\pm0.1$     & $1.34\pm0.08$   & \\
  $2^+$   & $B_2^{++}$ & $1.27\pm0.07$   
          & $1.3\pm0.1$     & $1.4\pm0.1$     & $1.3\pm0.1$ \\
  \hline
  \end{tabular}\end{center}
  \caption[dummy]{$SU(3)$ screening masses for $T>T_c$ on small
     ($4\times8^2\times16$), large ($4\times12^2\times16$) and
     short ($4\times8^2\times12$) lattices. The starred masses
     are obtained from two-mass fits, the rest from local masses.
     Data on $B_1^{-+}$ correlators was not collected in two cases.}
\label{tb.tc2}\end{table}

Our results for the screening masses are collected in Table \ref{tb.tc2}.
The most interesting result is the near equality of the $A_1^{++}$ and
$A_2^{-+}$ screening masses at $2T_c$. Similarly, the $B_1^{-+}$ and the
$B_2^{++}$ masses are equal and also equal to the $B_1^{++}$ mass.
Perturbation theory cannot be used to explain this pattern
of degeneracies because the $P=1$ correlators require two gluon exchange,
and the $P=-1$ correlation functions must have a minimum of four exchanged
gluons.

Finite volume effects are under good control, as shown by the three separate
runs on lattices of three sizes at $2T_c$. 
The study in \cite{old} had shown that finite lattice spacing effects are
also under control, by making two simulations at the same temperature but
at two different lattice spacings. The values of $m/T$ in the $A_1^{++}$
and $A_1^{-+}$ channels were seen to be independent of the lattice spacing.
We expect that this is true also of the screening masses in other channels,
but would certainly welcome a direct measurement.

A comparison with zero temperature results is simple because $T=0$
measurements have been performed at both these couplings
\cite{chen,forcrand,michael1}. Using the results in \cite{michael1} we
find that at $\beta=5.9$
\begin{equation}
   {m(3T_c/2,A_1^{++})\over m(T=0,A_1^{++})}\;=\;0.78\pm0.05,
      \qquad
   {m(3T_c/2,B_1^{++})\over m(T=0,E^{++})}\;=\;0.88\pm0.09.
\label{su3.comp1}\end{equation}
Moreover, at $\beta=6.0$ we find
\begin{equation}
   {m(2T_c,A_1^{++})\over m(T=0,A_1^{++})}\;=\;0.90\pm0.05,
      \qquad
   {m(2T_c,B_1^{++})\over m(T=0,E^{++})}\;=\;1.12\pm0.06.
\label{su3.comp2}\end{equation}
These comparisons are made between channels at $T=0$ which overlap the
corresponding channel at $T>0$. Although at this larger temperature the
$T=0$ $A_1^{++}$ mass is nearly equal to the thermal $A_1^{++}$ screening
mass, the state is completely different. The eigenvector of the variational
problem has equal overlaps with the $T=0$ $A_1^{++}$ and $E^{++}$ states.
It is clear that the physics observed here is completely different from the
$T=0$ physics.

\section{\label{sc.su2}$SU(2)$ Pure Gauge Theory}

The $SU(2)$ pure gauge theory with Wilson action was simulated at three
temperatures. With $N_\tau=4$ the critical coupling is \cite{fing2}
$\beta_c(N_\tau=4)\;=\;2.2998$. We performed two simulations close to
$T_c$--- one with $\beta=2.30$, and another at $\beta=2.25$. With the
lattice size we used, this other coupling is still within the critical
region, as indicated by the Polyakov loop susceptibility \cite{fing2}. We
also performed simulations at $T=2T_c$ with $\beta_c(N_\tau=8)=2.51$ and
at $T=4T_c$ with $\beta_c(N_\tau=16)=2.74$ \cite{fing1}.

These simulations were performed with an over-relaxation \cite{creutz} and a
Kennedy-Pendleton heat-bath algorithm \cite{kphb}. The class of loop operators
measured is listed in Appendix \ref{sc.irreps}. Each measurement of correlation
functions was separated by about one integrated auto-correlation time measured
through the Polyakov loop. The procedure for the analysis was identical to
that for $SU(3)$.

\subsection{\label{sc.tcsu2}$T\approx T_c$}

\begin{table}[htb]\begin{center}
  \begin{tabular}{|c|l|c|c|c|c|c|c|}  \hline
  $\beta$ & Operator & \multicolumn{3}{c|}{$(0,1)$}
                     & \multicolumn{3}{c|}{$(0,2)$}\\
  \cline{3-8} 
  & & range & $\chi^2$ & $\mu_0$ 
    & range & $\chi^2$ & $\mu_0$ \\ 
  \hline
  2.25 & $\left.A_1^+\right|_{T(A)}$ & [0:8] & 0.32 & $0.48^{+0.05}_{-0.06}$ 
                                     & [0:7] & 0.11 & $0.50^{+0.05}_{-0.06}$\\
  & $\left.A_1^+\right|_{T(E)}$ & [0:8] & 0.87 & $0.54^{+0.05}_{-0.07}$ 
                                & [0:8] & 0.26 & $0.56^{+0.06}_{-0.06}$ \\
  & $\left.A_1^+\right|_T$ & [0:8] & 1.08 & $0.53^{+0.05}_{-0.05}$ 
                           & [0:8] & 1.16 & $0.54^{+0.05}_{-0.05}$ \\
  & $\left.B_1^+\right|_{T(E)}$ & [0:5] & 0.08 & $1.26^{+0.08}_{-0.08}$ 
                                & [0:5] & 0.16 & $1.11^{+0.10}_{-0.10}$\\
  \hline
  2.30 & $\left.A_1^+\right|_{T(A)}$ & [0:8] & 0.50 & $0.31^{+0.01}_{-0.01}$
                                     & [0:8] & 1.02 & $0.32^{+0.01}_{-0.02}$ \\
  & $\left.A_1^+\right|_{T(E)}$ & [0:8] & 0.63 & $0.30^{+0.01}_{-0.01}$
                                & [0:8] & 0.55 & $0.30^{+0.01}_{-0.01}$ \\
  & $\left.A_1^+\right|_T$ & [0:8] & 0.40 & $0.30^{+0.01}_{-0.01}$
                           & [0:8] & 0.36 & $0.30^{+0.01}_{-0.01}$ \\
  & $\left.B_1^+\right|_{T(E)}$ & [0:5] & 0.50 & $1.10^{+0.04}_{-0.04}$
                                & [0:5] & 0.60 & $1.09^{+0.04}_{-0.04}$ \\
  \hline
  \end{tabular}\end{center}
  \caption{Screening masses in the $SU(2)$ theory near $T_c$.}
\label{tb.su2tc}\end{table}

\begin{figure}
\vskip10truecm
\includegraphics{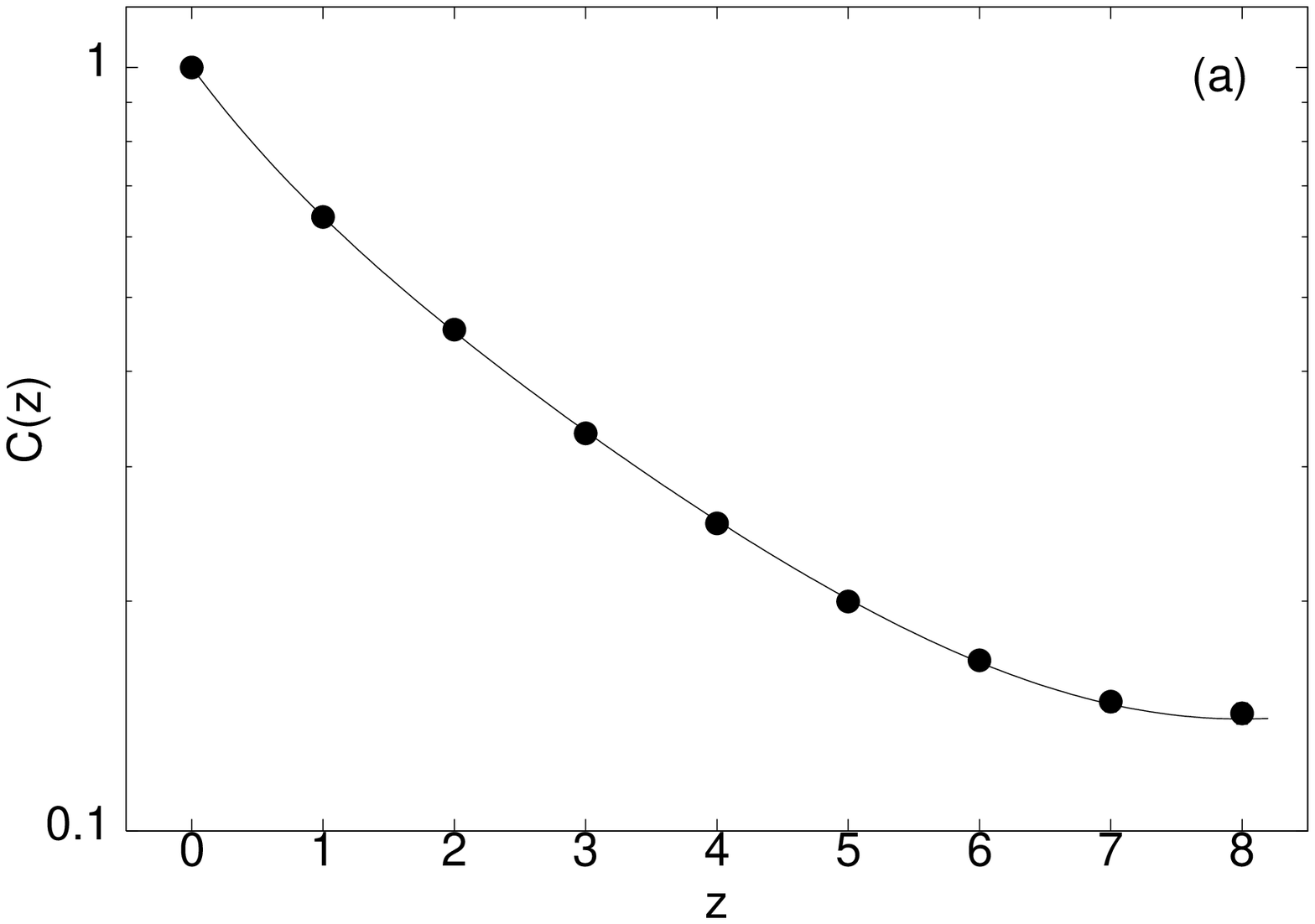}
\includegraphics{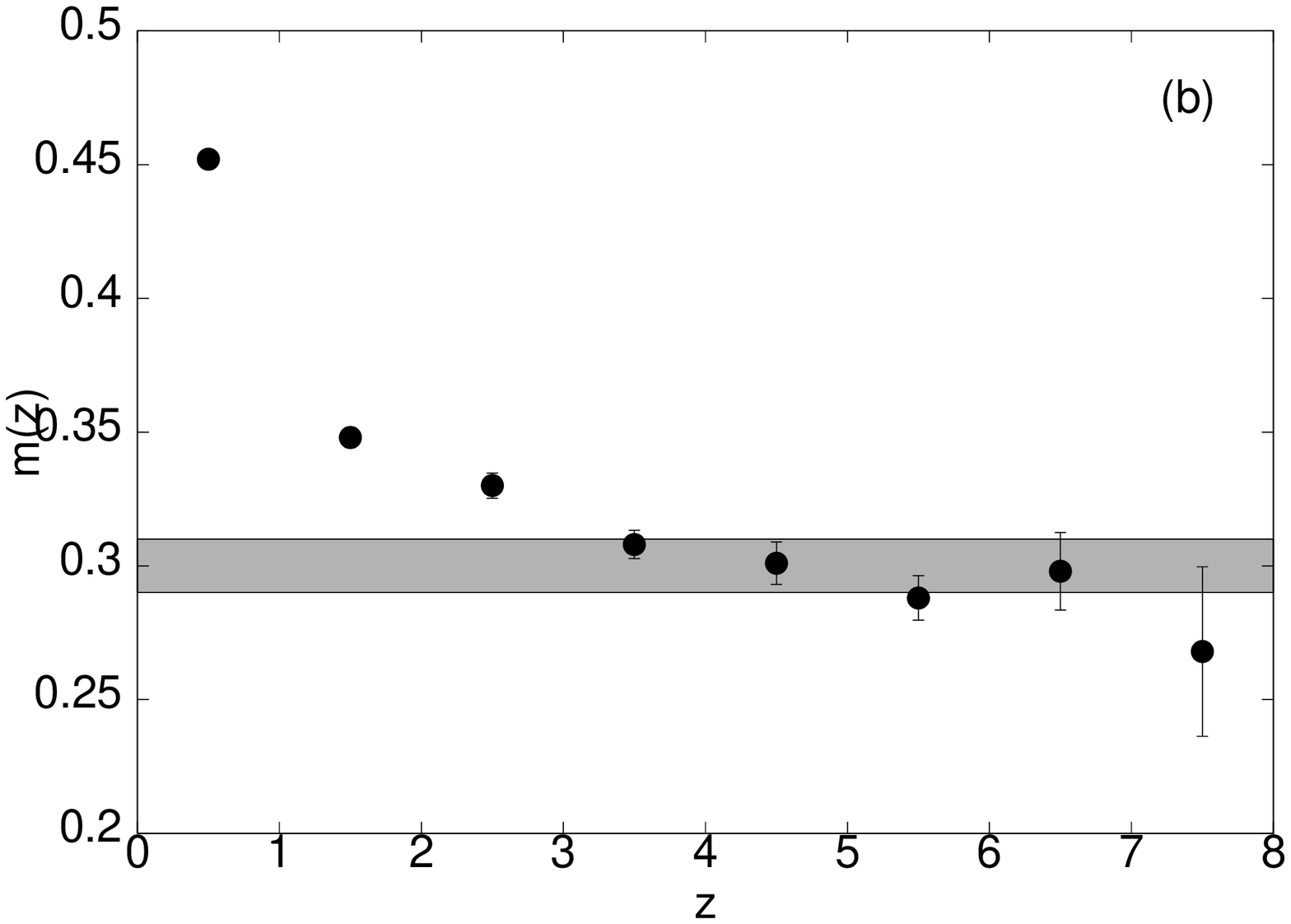}
\includegraphics{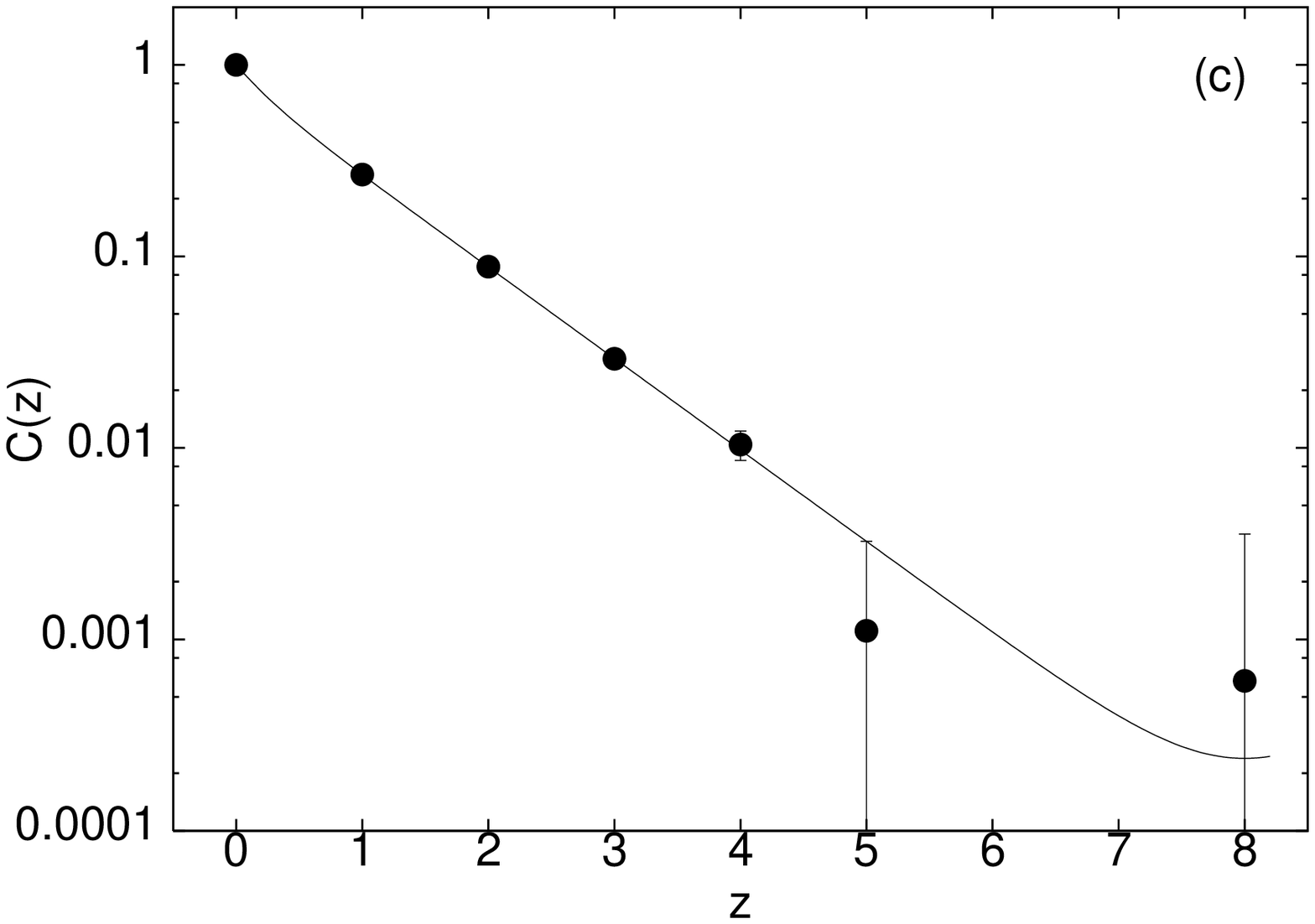}
\includegraphics{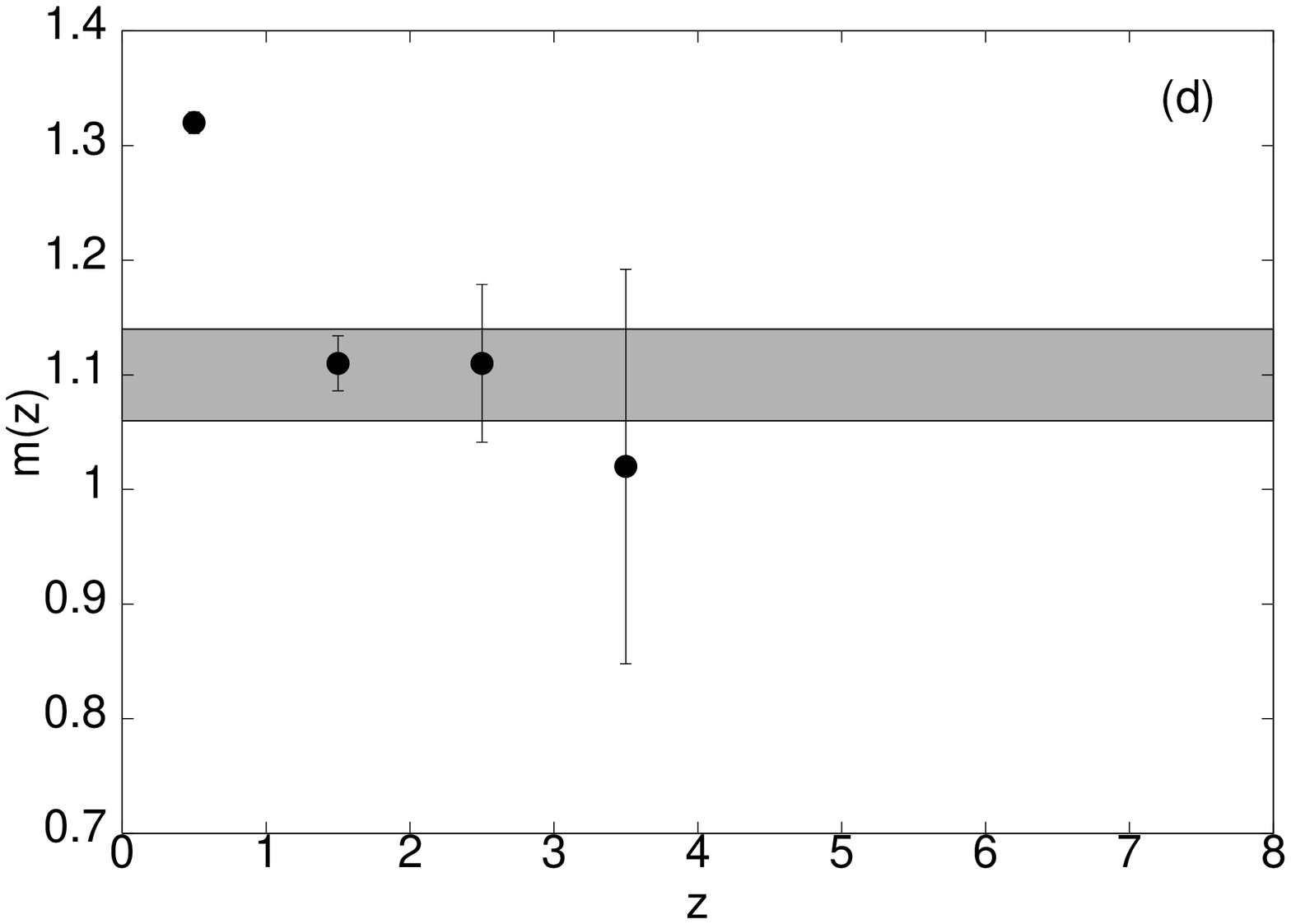}
\caption[dummy]{Correlation functions in $SU(2)$ at $\beta=2.3$ with $(0,1)$
   variation. (a) Data and fits for the thermal $A_1^+$ correlator, (b)
   $A_1^+$ local masses compared to the 1-$\sigma$ band around the best
   fit, (c) data and fits to the $B_1^+$ correlator, and (d) $B_1^+$
   local masses compared to the 1-$\sigma$ error band on the fit.}
\label{fg.su2tc}\end{figure}

For $\beta=2.30$, we took one measurement on a $4\times12^2\times16$ lattice
every 50 heatbath sweeps, and worked with $10000$ measurements after 
discarding the first 5000 sweeps for thermalisation. The set of operators 
measured is given in Appendix \ref{sc.irreps}. Our results for masses are 
presented in Table \ref{tb.su2tc} and the correlation functions and masses 
are displayed in Figure \ref{fg.su2tc}.

The $\left.A_1^+\right|_T$ masses coming from the $O_h$ irreps $A_1^+$ and
$E^+$ turn out to be identical. Note also that the parameter $r$ is rather
large, indicating a successful projection onto the ground state. The $A_1^+$
mass is significantly smaller than the $T=0$ mass at the same coupling,
$1.22\pm0.03$ \cite{michael2}.

The $B_1^+$ correlator is significantly more noisy. However the correlation
could be followed out to distance five. The mass estimate is fairly stable,
and the overlap with the ground state is rather good. This screening mass is
significantly higher than the $\left.A_1^+\right|_T$ screening mass, and
much smaller than $m(T=0,E^+)=1.94\pm0.08$ \cite{michael2}, from which it
comes.

At $\beta=2.25$ we took $10000$ measurements separated by 50 sweeps after
discarding the first 5000 sweeps. Our fits indicate that the two screening
masses for $\left.A_1^+\right|_T$, those from the $O_h$ irreps $A_1^+$ and
$E^+$ are almost degenerate at this point, although the common mass is
larger than that measured at $\beta=2.30$. The $\left.B_1^+\right|_T$
correlator was noisier, but good enough to yield a dependable measurement
of the mass. This screening mass is clearly split from the
$\left.A_1^+\right|_T$ mass coming from the same $O_h$ irrep $E^+$.
Our results are summarised in Table \ref{tb.su2tc}, and are in accord with
the expectation in Figure \ref{fg.guess}.

\subsection{\label{sc.su2hi}$T=2T_c$ and $4T_c$}

We performed runs with $4\times8^2\times16$ and $4\times12^2\times16$
lattices at $\beta=2.51$ ($2T_c$) and $\beta=2.74$ ($4T_c$). Three to
five over-relaxation sweeps were followed by one heat-bath sweep. Since
autocorrelation times of Polyakov loops and plaquettes were seen to be
less than three such composite steps, we performed a measurement on
every fifth composite step. In all these runs, the system quickly
relaxed into one of the $Z_2$ symmetric minima of the free energy and
stayed there through the duration of the run.

As explained in Appendix \ref{sc.irreps}, we measured two sets of
operators. The larger set, A, included all the operators in the smaller
set, B. The marginal improvement in the measurement of screening masses
did not compensate for the longer CPU time spent in constructing the extra
operators. On the smaller lattice we measured only the set A. At $2T_c$
we made 10000 measurements on the small lattice and 10000 measurements of
the operator set B on the larger lattice. At $4T_c$ we took 5000
measurements on the small lattice, 20000 of the operator set B and 10000
of the operator set A on the large lattice.

In the best cases we could follow the correlation function out to distance
5, and the local mass $m(3/2)$ already belonged to a plateau. The $B_1^-$
and $B_2^+$ channels were noisy, and we could follow the correlator only to
distance 3. For these two channels as well, we quote $m(3/2)$ as our estimate
of the local mass. The $A_2^-$ channel was more noisy than for the $SU(3)$
theory. We were unable to make any measurement in this channel. Unlike the
behaviour noticed for $SU(3)$, the analysis from $(0,1)$ variation gave
results in agreement with $(0,2)$ variation. Hence we report our analysis
from $(0,1)$ variation.

\begin{table}[htb]\begin{center}
\begin{tabular}{|c|c|c|c|c|c|}
   \hline
   $D^4_h$ & \multicolumn{2}{c|}{$2T_c$}
                     & \multicolumn{3}{c|}{$4T_c$} \\
   \cline{2-6}
   irrep & small A & large B & small A & large A & large B \\
   \hline
   $A_1^+$ & $0.71\pm0.05$   & $0.69\pm0.04$
           & $0.74\pm0.07$   & $0.73\pm0.05$ & $0.73\pm0.05$ \\
   $A_1^-$ & $1.14\pm0.02$   & $1.02\pm0.02$
           & $1.10\pm0.05$   & $0.87\pm0.04$ & $0.88\pm0.04$ \\
   $B_1^+$ & $1.18\pm0.03$   & $1.62\pm0.06$
           & $0.69\pm0.02^*$ & $1.08\pm0.06$ & $1.10\pm0.06$ \\
   $B_1^-$ & $1.9\pm0.2$     & $1.8\pm0.1$
           & $1.8\pm0.1$     & $1.55\pm0.06$ & $1.54\pm0.05$ \\
   $B_2^+$ & $1.8\pm0.1$     & $1.75\pm0.05$
           & $1.5\pm0.2$     & $1.60\pm0.08$ & $1.56\pm0.04$ \\
   \hline
\end{tabular}\end{center}
  \caption{Estimates of screening masses in the $SU(2)$ pure gauge theory
     from small ($4\times8^2\times16$) and large ($4\times12^2\times16$)
     lattices. The estimates are obtained from local masses, except the
     ones with a star. The latter are obtained from a fit.}
\label{tb.su2hi}\end{table}

The eigenvectors corresponding to the ground states are very stable in every
channel other than the $B_2^+$, where it showed large bin-to-bin fluctuations.
The eigenvalues were always more stable than the eigenvectors. The ground state
obtained by the variational method changed little between $2T_c$ and $4T_c$ for
the $A_1^+$, $A_1^-$ and $B_1^-$ states; the eigenvectors had an overlap
consistent with unity. However, in the $B_1^+$ channel, this overlap was
$0.70\pm0.04$. The screening mass $m(B_1^+)$ changes the most between $2T_c$
and $4T_c$.

Our results for the screening masses are summarized in Table \ref{tb.su2hi}.
The measurement of the screening mass in the $A_1^+$ sector gives
\begin{equation}
   {m(T,A_1^+)\over T}=\cases{
             2.8\pm0.2 & ($T=2T_c$),\cr
             2.9\pm0.2 & ($T=4T_c$).\cr}
\label{arat}\end{equation}
This is consistent with previous measurements from correlations of Polyakov
loops, reported in \cite{satz,pierre}. The near-equality of the $B_1^-$ and
the $B_2^+$ screening masses are characteristic of a dimensionally reduced
theory, and cannot be explained in perturbation theory.

The major difference between the $SU(2)$ and $SU(3)$ theories seems to be
in the volume dependence of various screening masses. The $A_1^+$ mass is
the only one which seems to be volume independent. The $B_1^+$ screening
mass increases with volume, and the rest decrease. A more extensive study
is required to obtain the infinite volume limit of these masses. Only after
this is done can we say more about the nature of the dimensionally reduced
theory.

Our measurements of $m(T,A_1^+)$ can be compared to the $T=0$ glueball masses
measured at $\beta=2.5$ and $2.6$ \cite{michael2}. Interpolation gives
$m(T=0,A_1^+)=0.70\pm0.03$ for $\beta=2.51$.
The mass ratio
\begin{equation}
   {m(2T_c,A_1^+)\over m(T=0,A_1^+)}=0.99\pm0.07,
\label{su2.rat1}\end{equation}
is close to unity. Thermal effects are seen in the ground state--- the
$A_1^+$ eigenvector has a large projection on both the $T=0$ $A_1^+$ and
$E^+$ channels. 
Glueball masses at $\beta=2.74$ for $T=0$ can be obtained by interpolating
between the measurements at $\beta=2.7$ \cite{michael3} and $\beta=2.85$
\cite{ukqcd}. We estimate $m(T=0,A_1^+)=0.35\pm0.03$. Then
\begin{equation}
   {m(4T_c,A_1^+)\over m(T=0,A_1^+)}=2.1\pm0.2,
\label{su2.rat2}\end{equation}
indicating a large thermal shift. This shows that the physics of
screening masses is quite different from that of $T=0$ glueball masses.

\section{\label{sc.final}Summary}

In this section we summarise the results of our lattice measurements
and discuss the physics implied by it. We try to deduce some general
features of the dimensionally reduced theory by comparing these
results with what is known of three dimensional gauge theories.

First we gather together our conclusions. We have found that for
$T\ge T_c$ the spectrum of screening masses is completely consistent
with a cylindrical symmetry of the spatial transfer matrix ($D^4_h$
on the lattice). At temperatures of $T_c$ and above, there is no remnant
of the $T=0$ $O(3)$ rotational symmetry. The scalar representation of
the cylinder group ($A_1^{++}$ on the lattice) gives the lowest
screening mass at all temperatures in both $SU(2)$ and $SU(3)$ theories.

Near $T_c$ this lowest screening mass is very small. Since the $SU(2)$
gauge theory undergoes a second order deconfining transition, it is
expected that this screening mass should be precisely zero on infinite
volume systems. The small non-zero value we observe can be ascribed to
finite-size effects.

The $SU(3)$ gauge theory has a first order deconfining
transition. Finite-sized systems near a first order phase transition show
a small screening mass, which vanishes in the infinite volume limit, and
is related to tunnelings, and hence the surface tension \cite{tension},
between the phases which coexist at a first order transition. Previous
measurements \cite{brown} had seen only this mass. A technical point of
interest to specialists is that our measurement of this mass yields
results consistent with previous observations.

The important quantity for the study of screening masses is not this, but the
finite screening mass--- the ``physical mass'' in the $A_1^{++}$ channel
for $SU(3)$ gauge theory. We have estimated it near $T_c$ for the first time.

\begin{table}[hbt]\begin{center}
  \begin{tabular}{|c|l|l|l|l|}  \hline
  $T$ & $A_1^{++}$ & $A_1^{-+}$ & $B_1^{++}$ & $B_2^{++}$ \\
      & $0_+^+$    & $0_-^+$    & $2^+$      & $2^+$ \\
  \hline
  $T_c$    & 3.4(4)  & - & 4.9(3) & - \\
  $\frac{3}{2}T_c$ &
             2.56(4) & 6.3(1) & 4.9(3) & 5.1(3) \\
  $2T_c$   & 2.60(4) & 6.3(2) & 4.8(1) & 5.6(4) \\
  \hline
  \end{tabular}\end{center}
  \caption{The ratio $m/T$ in various channels as a function of $T$ in the
     $SU(3)$ gauge theory. The $A_1^{++}$ screening mass shown near $T_c$ is
     the ``physical mass''. The masses are labelled by the name of the 4-d
     lattice irrep (upper line) and the 3-d continuum irrep (lower line).}
\label{tb.summary}\end{table}

Our results for the screening masses are summarised in Table \ref{tb.summary}.
As discussed in appendix \ref{sc.clebsch}, the spectrum is not
amenable to an understanding in terms of perturbative multigluon
states, and gives nonperturbative information about the underlying
effective theory.

Evidence for dimensional reduction at $2T_c$ comes from degeneracies in
the spectrum of the transfer matrix. In the $SU(3)$ theory $m(A_1^{++})
\simeq m(A_1^{-+})$, and $m(B_1^{-+})\simeq m(B_2^{++})\simeq m(B_1^{++})$.
The first two sets of equalities are sufficient to argue for dimensional
reduction of the lattice cutoff theory. The third equality implies $O(2)$
symmetry and hence allows us to make the stronger statement that the lattice
artifacts are small. In the $SU(2)$ theory, although we have not eliminated
lattice artifacts, dimensional reduction is shown by the relation
$m(B_2^+)\simeq m(B_1^-)$.

Now we turn to possible interpretations of our detailed observations.
The numerical values of the screening masses we have observed constrain
the form of the three dimensional effective theory that describes
equilibrium 4-d thermal gauge theories. In the scaling region of 3-d
$SU(N)$ pure gauge theories, the glueball mass ratios
\begin{equation}
   {m(A^{++})\over m(B^{++})}\;=\;0.60,\qquad
   {m(B^{++})\over m(A^{-+})}\;=\;0.78,
\label{final.rat}\end{equation}
are almost independent of $N$ for gauge groups $SU(N)$ \cite{2plus1}.
Our measurements at $2T_c$ in the $SU(3)$ theory give
\begin{equation}
   {m(A^{++})\over m(B^{++})}\;=\;0.54\pm0.02,\qquad
   {m(B^{++})\over m(A^{-+})}\;=\;0.76\pm0.02.
\label{final.ratour}\end{equation}
From the spectrum it seems likely that the dimensionally reduced theory
corresponding to the finite temperature $SU(3)$ theory may be a 3-d pure
gauge theory.

One final observation--- the $A_1^+$ mass in the $SU(2)$ theory agrees
with many other measurements (performed through Polyakov loop correlations),
and is expected to be independent of the lattice spacing. It is also seen
to be independent of the lattice volume. In addition, it agrees numerically
with the $A_1^{++}$ screening mass observed in the $SU(3)$ theory. It would
be interesting to study whether the infinite volume limit of the other
screening masses in both these theories show a similar agreement.

We will report on several technical points in future. A study of finite
lattice spacing effects is under way. We are also performing a more
extensive study of finite volume effects. We have not studied the
two-dimensional irreps, $E$, of $D^4_h$, but expect that the screening masses
in these channels would add to our understanding of this problem.

We would like to thank Rajiv Gavai for discussions.

\newpage\appendix

\section{\label{sc.irreps}Loop Operators}

We specify a loop with the notation $U_i(\mu,\nu,\lambda,\cdots)$. This
denotes a product of link matrices starting with $U_\mu(i)$ and proceeding
along the links in the directions $\nu$, $\lambda$, {\sl etc\/}.
The loops used in this work are drawn from the set---
\begin{eqnarray}
\label{8link}
\nonumber
   P^4 \;=\;& {\rm Re\ Tr\ }U_i(\mu,\nu,-\mu,-\nu),\qquad\qquad\qquad\\
\nonumber
   P^6 \;=\;& {\rm Re\ Tr\ }U_i(\mu,\mu,\nu,-\mu,-\mu,-\nu),\qquad\quad\\
\nonumber
   T^6 \;=\;& {\rm Re\ Tr\ }U_i(\mu,\nu,\rho,-\mu,-\nu,-\rho),\qquad\quad\\
\nonumber
   B^6 \;=\;& {\rm Re\ Tr\ }U_i(\mu,\nu,-\mu,\rho,-\nu,-\rho),\qquad\quad\\
\nonumber
   O^8_2 \;=\;& {\rm Re\ Tr\ }U_i(\mu,\mu,\nu,\rho,-\mu,-\mu,-\rho,-\nu),\\    
\nonumber
   O^8_3 \;=\;& {\rm Re\ Tr\ }U_i(\mu,\mu,\nu,-\mu,\nu,-\mu,-\nu,-\nu),\\
   O^8_4 \;=\;& {\rm Re\ Tr\ }U_i(\mu,\mu,\nu,-\mu,\rho,-\mu,-\rho,-\nu),\\
\nonumber 
   O^8_5 \;=\;& {\rm Re\ Tr\ }U_i(\mu,\mu,\nu,-\mu,\rho,-\nu,-\mu,-\rho),\\
\nonumber
   O^8_7 \;=\;& {\rm Re\ Tr\ }U_i(\mu,\mu,\nu,\rho,-\nu,-\mu,-\mu,-\rho),\\
\nonumber
   O^8_{14} \;=\;& {\rm Re\ Tr\ }U_i(\mu,\nu,-\mu,-\nu,-\mu,\rho,\mu,-\rho),\\
\nonumber
   O^8_{16} \;=\;& {\rm Re\ Tr\ }U_i(\mu,\nu,-\mu,-\nu,-\rho,-\nu,\rho,\nu),\\
\nonumber
   O^8_{17} \;=\;& {\rm Re\ Tr\ }U_i(\mu,\nu,-\mu,-\nu,-\rho,\nu,\rho,-\nu),\\
\nonumber
   O^8_{18} \;=\;& {\rm Re\ Tr\ }U_i(\mu,\mu,\nu,-\mu,\rho,-\mu,-\nu,-\rho).
\end{eqnarray}
Our convention for naming these loops follows that of \cite{bb}. The plaquette
($P^4$), 6-link planar ($P^6$), twisted ($T^6$) and bent ($B^6$) loops, and
the 8-link loop $O^8_{14}$ were considered in detail earlier \cite{old}. We
have also used the double traversal of some of these loops---
\begin{equation}
   O^2\;=\;{\rm Re\ Tr\ }U_i U_i,\qquad{\rm where}\qquad
   O\;=\;{\rm Re\ Tr\ }U_i,
\label{double}\end{equation}
where $U_i$ is the $SU(N)$ matrix corresponding to a loop. The representation
content of such pairs $O$ and $O^2$ are identical.

The irreps of the symmetry group can be constructed by acting on any loop
by the projection operators of the group. For $D^4_h$ we use the operators
\begin{eqnarray}
\nonumber
   A_1^\pm\;=&\;(E+C_4)(E+C_4^2)(E+C_2)(E\pm P),\\
\nonumber
   A_2^\pm\;=&\;(E+C_4)(E+C_4^2)(E-C_2)(E\pm P),\\
   B_1^\pm\;=&\;(E-C_4)(E+C_4^2)(E+C_2)(E\pm P),\\
\nonumber
   B_2^\pm\;=&\;(E-C_4)(E+C_4^2)(E-C_2)(E\pm P).
\label{irreps}\end{eqnarray}
We have not used the remaining four projectors, which are the two independent
sets each of $E^+$ and $E^-$ projectors that can be obtained by changing the
sign of $C_4^2$ in the above formul\ae. The irrep content of the loops in
eq.\ (\ref{8link}) can simply be obtained using these projectors in a small
Mathematica program.

In the $SU(3)$ measurements we used the following set of operators---
\[\begin{tabular}{ll}
   $A_1^{++}\quad$  & (2) $P^4$, (2) $(P^4)^2$, (3) $P^6$, (3) $(P^6)^2$,
                      (1) $O^8_{14}$, (1) $O^8_{16}$, (1) $O^8_{17}$,\\
                    & (3) $O^8_{18}$, (3) $(O^8_{18})^2$ \\
   $A_1^{-+}\quad$  & (1) $O^8_{14}$, (1) $O^8_{16}$,
                      (3) $O^8_{18}$, (3) $(O^8_{18})^2$ \\
   $A_2^{-+}\quad$  & (1) $B^6$, (1) $(B^6)^2$,
                      (1) $O^8_{14}$, (1) $O^8_{16}$, (1) $O^8_{17}$,
                      (3) $O^8_{18}$, (3) $(O^8_{18})^2$ \\
   $B_1^{++}\quad$  & (1) $P^4$, (1) $(P^4)^2$, (3) $P^6$, (3) $(P^6)^2$,
                      (1) $O^8_{14}$, (1) $O^8_{16}$,\\
                    & (3) $O^8_{18}$, (3) $(O^8_{18})^2$ \\
   $B_1^{-+}\quad$  & (1) $O^8_{14}$, (1) $O^8_{16}$, (3) $O^8_{18}$,
                      (3) $(O^8_{18})^2$ \\
   $B_2^{++}\quad$  & (1) $O^8_{14}$, (1) $O^8_{16}$,
                      (3) $O^8_{18}$, (3) $(O^8_{18})^2$ \\
\end{tabular}\]
For the simulation near $T_c$ only the plaquette and 6-link planar
loops were used.

For the $SU(2)$ measurements a bigger set of operators was used---
\[\begin{tabular}{ll}
   $A_1^{+}\quad$  & (2) $P^4$, (3) $P^6$, (2) $B^6$, (1) $T^6$,
		     (2) $O^8_2$, (2) $O^8_3$, (3) $O^8_4$, \\
                   & (2) $O^8_5$, (3) $O^8_7$, (2) $O^8_{14}$,
		     (2) $O^8_{16}$, (2) $O^8_{17}$, (3) $O^8_{18}$ \\
   $A_1^{-}\quad$  & (3) $O^8_4$, (2) $O^8_5$, (2) $O^8_{14}$,
                     (2) $O^8_{16}$, (3) $O^8_{18}$ \\
   $A_2^{+}\quad$  & (3) $O^8_4$, (1) $O^8_5$, (1) $O^8_7$,
                     (1) $O^8_{14}$, (1) $O^8_{16}$,(3) $O^8_{18}$ \\
   $A_2^{-}\quad$  & (1) $B^6$, (1) $O^8_2$, (1) $O^8_3$, (3) $O^8_4$,
                     (1) $O^8_5$, (2) $O^8_7$, (1) $O^8_{14}$, \\
                   & (1) $O^8_{16}$, (1) $O^8_{17}$, (3) $O^8_{18}$ \\
   $B_1^{+}\quad$  & (1) $P^4$, (3) $P^6$, (1) $B^6$, (1) $O^8_2$,
		     (1) $O^8_3$, (3) $O^8_4$, (1) $O^8_5$, \\
		   & (3) $O^8_7$, (1) $O^8_{14}$, (1) $O^8_{16}$,
                     (1) $O^8_{17}$, (3) $O^8_{18}$ \\
   $B_1^{-}\quad$  & (3) $O^8_4$, (1) $O^8_5$, (1) $O^8_{14}$,
                     (1) $O^8_{16}$, (3) $O^8_{18}$ \\
   $B_2^{+}\quad$  & (1) $B^6$, (1) $T^6$, (1) $O^8_2$, (1) $O^8_3$,
                     (3) $O^8_4$, (2) $O^8_5$, (1) $O^8_7$, \\
                   & (2) $O^8_{14}$, (2) $O^8_{16}$,
                     (1) $O^8_{17}$, (3) $O^8_{18}$ \\
   $B_2^{-}\quad$  & (1) $B^6$, (1) $O^8_2$, (1) $O^8_3$, (3) $O^8_4$,
                     (2) $O^8_5$, (2) $O^8_7$, (2) $O^8_{14}$, \\
                   & (2) $O^8_{16}$, (1) $O^8_{17}$, (3) $O^8_{18}$ \\
\end{tabular}\]
along with the double traversals of each of them. This full set is the
SET A of Section \ref{sc.su2hi}. Set B contained only the operators
$P^4$, $P^6$, $B^6$, $T^6$, $O^8_4$, $O^8_{14}$, $O^8_{16}$ and
$O^8_{18}$. For the simulations near $T_c$ only the plaquette and the
six-link loops were used.

\section{\label{sc.clebsch}Multi-gluon States}

The simplest way of understanding the screening masses in the high 
temperature phase would be to accomodate it in a perturbative 
framework. Correlators of loops will then be dominated by suitable
multigluon states, and the screening masses would give information
on $M_D$ and $M_m$ \cite{old}. Fortunately, such a hypothesis is open
to a direct numerical test. In this appendix we show that the screening
masses cannot be understood in perturbation theory.
 
\begin{table}[bht]\begin{center}
   \begin{tabular}{|c|l|l|}
   \hline
   $\bf k$ & $A_t$ & $A_x, A_y$ \\
   \hline

   (0,0,0) & $A_2^-$
           & $E^-$ \\

   (0,k,0) & $A_2^-, B_2^-, E^+$
           & $A_1^+, B_1^+, E^-$ \\

   (k,0,0) & $A_1^+, A_2^-$
           & $E^+, E^-$ \\

   (0,k,k) & $A_2^-, B_1^-, E^+$
           & $A_1^+, A_2^+, B_1^+, B_2^+, 2 E^-$ \\

  (0,k,k') & $A_1^-, A_2^-, B_1^-, B_2^-, 2 E^+$
           & $A_1^+, A_2^+, B_1^+, B_2^+, 2 E^-$ \\

  (k,k',0) & $A_1^+, A_2^-, B_1^+, B_2^-, E^+, E^-$
           & $A_1^+, A_2^-, B_1^+, B_2^-, E^+, E^-$ \\

  (k,k',k'') & $A_1^\pm, A_2^\pm, B_1^\pm, B_2^\pm, E^\pm, E^\pm$
             & $A_1^\pm, A_2^\pm, B_1^\pm, B_2^\pm, E^\pm, E^\pm$ \\
  \hline
  \end{tabular}\end{center}
  \caption[dummy]{The $D^4_h$ representation content of gluon operators
     at various points, $\bf k$, of the Brillouin zone. The first
     component of $\bf k$ is $k_0$. Note that all these irreps
     come with $C=-1$.}
\label{tb.glrep}\end{table}

We first classify the irreps of $D^4_h$ which can be obtained by specific
multi-gluon exchanges. The gluons carry arbitrary
momenta, and are classified by representations of the space group.
We break these irreps of the space group under the point group.
Then the allowed representations for colour singlet zero momentum
correlators are obtained by combining the gluon representations by
the Clebsch-Gordan series for the point group $D^4_h$ and applying
appropriate exchange symmetries. 

We begin by specifying the lattice analogue of gluon field operators in
momentum space by the Fourier transform of a projection of link matrices
onto the $SU(N)$ algebra---
\begin{equation}
   A_\mu({\bf k})\;=\;i\sum_{\bf x}{\rm e}^{i{\bf k}\cdot{\bf x}}
       \bigl[U_\mu({\bf x})-U^\dagger_\mu({\bf x})
                  - {\rm Im\,Tr}\,U_\mu({\bf x}) \bigr].
\label{gluon}\end{equation}
Here $\bf x$ takes values in a slice orthogonal to the $z$-direction,
and $\bf k$ in the
corresponding Brillouin zone. In general, the action of $D^4_h$ carries
one $\bf k$ into another. The representations built over the orbit of
$\bf k$ under the action of $D^4_h$ are usually large and reducible.
The representation content of such gluon field operators is gauge
independent, and is shown in Table \ref{tb.glrep}.

\begin{table}[bht]\begin{center}
   \begin{tabular}{|c|l|l|l|l|}\hline
   $\bf k$ & $G^{(2)}_{tt}$
           & $G^{(2)}_{tx}, G^{(2)}_{ty}$
           & $G^{(2)}_{xx}, G^{(2)}_{yy}$
           & $G^{(2)}_{xy}$ \\
   \hline

   (0,0,0) & $A_1^+$
           & $E^+$
           & $A_1^+, B_1^+$
           & $B_2^+$ \\

   (0,k,0) & $A_1^+, B_1^+$
           & $E^+$
           & $A_1^+, B_1^+$
           & $A_2^+, B_2^+$ \\

   (k,0,0) & $A_1^+$
           & $E^+$
           & $A_1^+, B_1^+$
           & $B_2^+$ \\

   (0,k,k) & $A_1^+, B_2^+$
           & $2 E^+$
           & $A_1^+, A_2^+,$
           & $A_1^+, B_2^+$ \\

           & 
           & 
           & $B_1^+, B_2^+$
           & \\

  (0,k,k') & $A_1^+, A_2^+,$
           & $2 E^+$
           & $A_1^+, A_2^+,$
           & $A_1^+, A_2^+,$ \\

           & $B_1^+, B_2^+$
           & 
           & $B_1^+, B_2^+$
           & $B_1^+, B_2^+$ \\

  (k,k',0) & $A_1^+, B_1^+, E^+$
           & $A_1^+, B_1^+, E^+$
           & $A_1^+, B_1^+, E^+$
           & $A_2^+, B_2^+, E^+$ \\

  (k,k',k'') & $A_1^+, A_2^+, B_1^+,$
             & $A_1^+, A_2^+, B_1^+,$
             & $A_1^+, A_2^+, B_1^+,$
             & $A_1^+, A_2^+, B_1^+,$ \\

             & $B_2^+, E^+$
             & $B_2^+, E^+$
             & $B_2^+, E^+$
             & $B_2^+, E^+$ \\
  \hline
  \end{tabular}\end{center}
  \caption[dummy]{The irreps for symmetric colour singlet $C=1$
     two-gluon states with total momentum zero are listed for
     the momentum, $\bf k$, of one of the gluon at various points
     in the Brillouin zone.}
\label{tb.clebsch}\end{table}

Gauge invariant $C=1$ states of zero momentum can be constructed as linear
combinations of the composite operators
\begin{equation}
   G^{(2)}_{\mu\nu}\;=\;Tr\,\bigl[A_\mu({\bf k}) A_\nu(-{\bf k})\bigr],
\label{ball}\end{equation}
where the trace is over $SU(N)$ generators. Correlators of such an 
operator will decay, in leading order, with the mass
$E_\mu({\bf k})+E_\nu({\bf k})$, where $E_0=E_D$, $E_{1,2}=E_m$ and 
\begin{equation}
   \sinh^2 (E_{D,m} ({\bf k}) / 2)\;=\;\sinh^2 (M_{D,m} / 2) 
       + \sum_{i=0}^2 \sin^2 (k_i / 2).
\label{massreln}\end{equation}
The cyclic property of traces
ensures that $G^{(2)}$ is symmetric under any operation that flips
the polarisation indices and simultaneously changes the sign of $\bf k$.
In addition, if we require that the state be symmetric under the exchange
of the gluon fields, then only $P=1$ irreps are allowed. The representation
content of these operators in all parts of the Brillouin zone is given in
Table \ref{tb.clebsch}.

In order to obtain $P=-1$, $C=1$ irreps we have to go to combinations
of four gluon field operators. Colour singlet gauge invariant correlators
with $C=-1$ start with composite operators of three gluon fields.

The reduction  of multi-gluon operators then tells us that
\begin{itemize}
\item
   For $C=1$, the lowest mass observed in any $P=1$ channel would be at
   most half of the lowest mass observed in any $P=-1$ channel, since the
   former are obtained by two-gluon exchange but the latter require at
   least four-gluon exchange.
\item
   If $E_m({\bf 0}) < E_D({\bf 0})$, then 
   $m(A_1^{++})=m(B_1^{++})=m(B_2^{++})$.
\item
   If $E_D({\bf 0}) < E_m({\bf 0}) < E_D({\bf K})$, for ${\bf K}=(0,
   2\pi/N_x,0)$, then $m(B_1^{++})= m(B_2^{++})$ and the mass difference
   between the $A_1^{++}$ and $B_1^{++}$ states is $2[E_m({\bf 0})-
   E_D({\bf 0})]$. Here $N_x$ is the spatial size of the $z$-slice.
\item
   If $E_m({\bf 0}) > E_D({\bf K})$, then $m(B_1^{++}) < m(B_2^{++})$ and the
   mass difference $m(B_1^{++}) - m(A_1^{++})$ decreases with lattice size.
\end{itemize}
The first condition is clearly violated in the spectra we obtain for
both $SU(2)$ and $SU(3)$ theories. At $2T_c$ in both the theories
we found $m(B_1^{-+})\simeq m(B_2^{++})$. In addition, in the $SU(3)$
theory $m(A_1^{++})\simeq m(A_2^{-+})$. Both these observations violate
the first condition. This is sufficient evidence for a failure of 
the interpretation of these masses as screening masses for
perturbative multigluon states.

If we ignore this, and restrict ourselves to the $P=1$, $C=1$ states
only, then one might be tempted to match the $SU(3)$ spectrum to the
third condition above, since the mass difference 
$m(B_1^{++})-m(A_1^{++})$ is non-zero and independent of the lattice 
size. However, from eq.\ (\ref{massreln}), the condition 
$M_m < E_D ({\bf k})$ is satisfied for $N_x=8$ but not for 
$N_x=12$. Thus, none of the conditions above hold, and we conclude
that the spectrum of screening masses cannot be given a perturbative
interpretation of multigluon states, and indeed gives information
about the nonperturbative spectrum of the effective theory. A similar 
statement has been made in ref. \cite{brn} in the context of the 
massgap obtained from Polyakov loop correlators at moderate temperatures,
on the basis of the detailed decay patterns of the correlators seen in 
lattice studies.

\section{\label{sc.fuzz}Improved Operators}

Loop correlations are known to be very noisy. In order to increase the
signal/background ratio, we used a hybrid of Teper's doubled-link fuzzing
procedure \cite{teper1} and the smearing procedure adopted by the APE
collaboration \cite{ape1}. We define fuzzed links at level $l+1$ recursively
in terms of those at level $l$ by the equation
\begin{equation}
   {\rm max}\,{\rm tr}\left(M^\dagger_{i,\mu}U^{(l+1)}_{i,\mu}\right)
   \quad{\rm where}\quad
   M_{i,\mu}\;=\; U^{(l)}_{i,\mu} + \sum_{\nu\ne\mu,\hat z}
     U^{(l)}_{i,\nu} U^{(l)}_{i+\nu,\mu} U^{\dagger(l)}_{i+\mu,\nu},
\label{fuzz}\end{equation}
$U^{(l+1)}$ are elements of $SU(N)$, and the links for $l=0$ are those
generated by the Monte Carlo procedure. For general $SU(N)$, this
maximisation is most easily accomplished using the ``polar'' decomposition
of a general complex matrix to write
\begin{equation}
   M\;=\;\omega U^{(l+1)}H,
\label{proj}\end{equation}
where $\omega$ is a complex number of unit modulus, $H$ is hermitian
and $U^{(l+1)}$ is special unitary. There is a discrete ambiguity in
this decomposition, corresponding to the signs of the eigenvalues
of $H$. When all the eigenvalues of $H$ are chosen to be positive,
$U^{(l+1)}$ maximises the trace in eq.\ (\ref{fuzz})\footnote{We would
like to thank Gautam Mandal and Avinash Dhar for a discussion of this
point.}. For $SU(2)$ the
algorithm is simpler since $H$ is a multiple of the identity. The
projection then involves only a division of $M$ by the square root
of its determinant.

In a test run with $SU(3)$ at $\beta=5.7$ on a $4^3\times12$ lattice, the
procedure in eq.\ (\ref{fuzz}) was found to perform better than doubled-link
fuzzing. Since the latter technique is known to work well on larger lattices,
we conclude that the problem is due to the fact that with small lattices, only
a small number of doubled-link fuzzing steps is possible. Presumably on larger
lattices, where more fuzzing levels can be reached, equally good results can
be obtained with either fuzzing technique. For the $SU(3)$ theory we worked
with eq.\ (\ref{fuzz}) and $l\le4$.

For $SU(2)$, since we use a lattice which has 12 spatial sites, upto three
levels of doubled-link fuzzing can be performed for the spatial links. For
finite temperature problems it is perfectly all right if the number of fuzzing
steps in the time direction is different from that in other directions. We
perform only one doubled-link fuzzing in the time direction. We
checked that using three levels of doubled-link fuzzing gave a better projection
than seven steps of (C.1), and therefore used the former, for our runs near
$T_c$. For our runs at $2 T_c$ and $4 T_c$, we experimented with a combination
of doubled-link fuzzing and (C.1); using a combination of one doubled-link
fuzzing followed by two steps of (C.1), the whole set being repeated once.
Since this gave a slight improvement over three steps of doubled-link fuzzing,
we used this technique at these higher temperatures. For our runs on the
smaller lattice we used one doubled-link fuzzing followed by 4 steps of
eq.\ (\ref{fuzz}).

\section{\label{sc.vary}Variational Correlators}

It is not known a priori which linear combination of loop operators
acting on the vacuum generates the state with the lowest mass in a channel
with given quantum numbers. However, such a state will give a correlation
function which has the slowest possible decay with increasing separation.
Given a basis set of loop operators, we can try to construct a linear
combination which satisfies this property of slowest decay. This is
the idea of a widely used variational technique \cite{proj}. Since we have
found no discussion in the literature of a numerically stable algorithm for
its implementation, we document such a method here.

We construct cross correlations between all the loop operators at our
disposal to yield the (symmetric) matrix of correlations $C_{ij}(z)$.
A combination of operators which has large projection to the ground state
is obtained by solving the variational problem over $Y$---
\begin{equation}
  {\rm max}\left[{Y^TC(z_1)Y\over Y^TC(z_0)Y}\right]=
      \lambda(z_0,z_1),\qquad(z_0<z_1).
\label{vary}\end{equation}
If $C(z_0)$ is positive definite, as guaranteed by the reflection positivity
of the Wilson action, then this extremisation problem reduces to finding the
maximum eigenvalue of the system---
\begin{equation}
   C(z_1)Y\;=\;\lambda(z_0,z_1)C(z_0)Y.
\label{geig}\end{equation}
With the corresponding eigenvector, we define the $(z_0,z_1)$ variational
correlator
\begin{equation}
   \widetilde C_{z_0,z_1}(z)\;=\;Y^T C(z)Y.
\label{corr}\end{equation}
We can utilise the freedom of normalising the eigenvector $Y$ to set the
variational correlator to unity at separation $z=0$.

Reflection positivity of the action guarantees that $C(z_0)$ is positive
definite. It can be treated as a metric, and after appropriate scaling, the
problem in eq.\ (\ref{vary}) can be phrased as the extremisation of a
quadratic form over a sphere--- leading to the usual matrix eigenvalue
problem \cite{matrix}. Algorithmically, this naive idea can be
implemented by transforming both sides of eq.\ (\ref{geig}) to the basis
where $C(z_0)$ is diagonal, absorbing the diagonal elements into $Y$ by
appropriate rescaling, and then solving the usual eigenvalue problem for
this transformed $C(z_1)$. However, if some of the eigenvalues of $C(z_0)$
are small, then the extremum problem is ill-conditioned because the solution
is sent off to infinity along the nearly flat directions.

With finite statistics the problem may be even worse. Due to statistical
fluctuations, the measured correlation matrix may not be positive definite.
It is then better to treat the problematic directions as exactly flat, since
this discards the subset of the data which is most corrupted by noise. The
solution is easily specified by going to the basis in which $C(z_0)$ is
diagonal and blocking the matrices into the form
\begin{equation}
  C(z_1)=\left(\matrix{C_{11}&C_{12}\cr C^T_{12}&C_{22}}\right),
    \qquad{\rm and}\qquad
  C(z_0)=\left(\matrix{G&0\cr0&0}\right),
\label{blocks}\end{equation}
where the eigenvalues, $e_i$, of $C(z_0)$ which satisfy the cut condition
\begin{equation}
    e_i\;<\;\epsilon e_0
\label{cut}\end{equation}
have been set to zero. Here $e_0$ is the maximum eigenvalue of $C(z_0)$.
The diagonal sub-matrix $G$ is positive definite and has a condition number
less than $1/\epsilon$. In this basis, eq.\ (\ref{vary}) is equivalent to
the set of equations---
\begin{equation}
   C_{11}Y_1 + C_{12}Y_2 = \lambda G Y_1
        \qquad{\rm and}\qquad
   C^T_{12}Y_1 + C_{22}Y_2 = 0.
\label{sys1}\end{equation}
Solving the latter for $Y_2$ and substituting into the former gives the
eigenvalue problem---
\begin{equation}
   \left[C_{11} - C_{12}C^{-1}_{22}C^T_{12}\right] Y_1 = \lambda G Y_1,
\label{sys2}\end{equation}
which is well-defined and numerically well-conditioned as long as $C_{22}$
is invertible.

Notice, however, that eq.\ (\ref{geig}) is ill-conditioned only if both
$C(z_1)$ and $C(z_0)$ have nearly flat directions; otherwise their roles
may be interchanged by inverting eq.\ (\ref{vary}) and converting the
maximum problem into that of finding a minimum. If this cannot be done,
then we must take care of the case that $C_{22}$ is not invertible. 

The $m\times n$ matrix $C^T_{12}$ is a map from $R^n$ to $R^m$
({\sl i.e.\/}, $Y_1$ has $n$ real components and $Y_2$ has $m$). Its range
is the subset of $R^m$ to which the whole of $R^n$ is mapped.
If $C_{22}$ is singular, then its null-space is contained in the complement
of the range of $C^T_{12}$. As a result, the second of eq.\ (\ref{sys1})
can be solved by a left-multiplication by any matrix which coincides with
the inverse of $C_{22}$ in the complement of its null-space ({\sl i.e.}, in
the range of $C^T_{12}$). Therefore, in eq.\ (\ref{sys2}) $C^{-1}_{22}$ can
be replaced by a pseudo-inverse
\begin{equation}
   C^{-1}_{22}=V^T\bar\Lambda^{-1} V \qquad{\rm where}\qquad
   C_{22}=V^T\Lambda V,
\end{equation}
$V$ is an orthogonal matrix, $\Lambda$ is diagonal, and the small
components of $\Lambda$ are set to zero in the pseudo-inverse
$\bar\Lambda^{-1}$. With this definition of $C_{22}^{-1}$, the
generalised eigenvalue problem is completely well-defined and
eq.\ (\ref{sys2}) may be solved by the naive algorithm described
earlier.

\begin{figure}
\vskip6truecm
\includegraphics{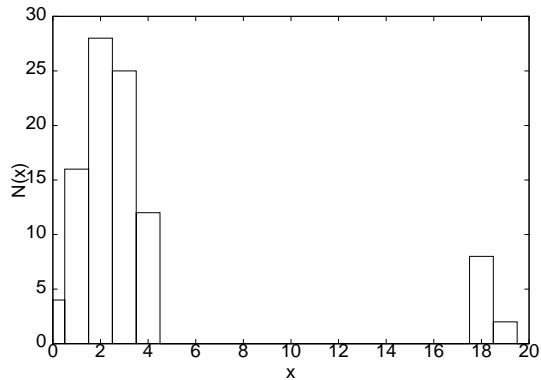}
\caption[dummy]{Histogram of normalised eigenvalues of the $95\times95$
    correlation matrix at distance zero, $C(0)$ in the $A_1^{++}$ channel
    at $2T_c$ for the $SU(3)$ theory.}
\label{fg.eigs}\end{figure}

A very nice numerical illustration is provided by our data on the
$95\times95$ matrix of the $A_1^{++}$ correlation function at $2T_c$ for the
$SU(3)$ gauge theory (section \ref{sc.su3hi}). The distribution of
\begin{equation}
   x\;=\;\log_{10}\left({e_0\over|e_i|}\right),
\label{example}\end{equation}
is shown in Figure \ref{fg.eigs}. The ten small eigenvalues clustered at the
end of a huge spectral gap have both positive and negative signs, and are
due to noise in the data. Because they are so well separated, the cut
$\epsilon$ (in eq.\ \ref{cut}) can be chosen to have any value between
$10^{-4}$ and $10^{-17}$. The results for eigenvalues and eigenvectors are
stable in this whole range of choices. Somewhat smaller values,
$\epsilon\approx10^{-3}$, are also found to be acceptable and are in fact
preferred for reasons of numerical stability of the linear algebra routines.
Removing the cut destabilises the problem completely. The spectral
distributions are similar in most channels for both $SU(2)$ and $SU(3)$
theories. In a few cases the spectral distribution is gapless but has a
long tail. Even in such cases, $\epsilon\approx10^{-3}$--$10^{-4}$ give
stable results.

\end{document}